\documentclass[aps,prb,twocolumn,superscriptaddress]{revtex4-1}
\usepackage{graphicx}
\usepackage{amsmath}
\usepackage{hyperref}
\pdfoutput=1 

\begin{document}

\title{Efficient Learning of a One-dimensional Density Functional Theory}

\author{M. Michael Denner}
\author{Mark H. Fischer}
\author{Titus Neupert}

\affiliation{Department of Physics, University of Zurich, Winterthurerstrasse 190, 8057 Zurich, Switzerland}

\date{\today}

\begin{abstract}

Density functional theory underlies the most successful and widely used numerical methods for electronic structure prediction of solids. However, it has the fundamental shortcoming that the universal density functional is unknown. In addition, the computational result---energy and charge density distribution of the ground-state---is useful for electronic properties of solids mostly when reduced to a band structure interpretation based on the Kohn-Sham approach.
Here, we demonstrate how machine learning algorithms can help to free density functional theory from these limitations. We study a theory of spinless fermions on a one-dimensional lattice. The density functional is implicitly represented by a neural network, which predicts, besides the ground-state energy and density distribution, density-density correlation functions. At no point do we require a band structure interpretation. 
The training data, obtained via exact diagonalization, feeds into a learning scheme inspired by active learning, which minimizes the computational costs for data generation.
We show that the network results are of high quantitative accuracy and, despite learning on random potentials, capture both symmetry-breaking and topological phase transitions correctly. 

\end{abstract}

\maketitle

\section{Introduction}
Materials with strong electronic correlations host a variety of intriguing phenomena and quantum phases. Modeling and understanding these systems are among the greatest challenges in theoretical condensed matter physics. For quantitative and predictive results, numerical calculations are indispensable. The most widely and successfully used numerical approach to the electronic structure problem is based on density functional theory (DFT). 
In condensed matter physics, DFT is often linked to band structure calculations, while it is in principle much more powerful than that. The Hohenberg-Kohn theorems guarantee that a (potentially correlated) many-body ground state is uniquely determined by its energy and charge density distribution~\cite{Hohenberg:1964}. However, for practical implementations and a physical interpretation of calculated results, the Kohn-Sham ansatz is commonly used, producing the band structure of a different, non-interacting system with the same energy and density~\cite{Kohn:1965}. The implicit assumption is that this band structure captures the essential physics of the original system, at least if correlations are weak enough. 

A critical shortcoming of DFT is that its eponymous functional is not known; instead, approximations on various levels of complexity are commonly employed~\cite{Mori-sanchez:2008}. It is important to emphasize that most of the functional is \emph{universal}, representing the many-particle Schr{\"o}dinger equation. The only nonuniversal input in a DFT calculation for a crystal is the potential landscape within the unit cell induced from the ions and core electrons as well as the particle number, both of which do not affect the universal part of the functional.

The recent rise of machine-learning methods used to model physical systems has sparked hopes to use these methods for improving DFT calculations~\cite{Schleder:2019}. The approaches interject the DFT workflow at various stages, ranging from improving the Kohn-Sham scheme by representing the exchange-correlation functionals~\cite{Lundgaard:2016, Kolb:2017, Liu:2017, Nagai:2018, Dick:2019, Schmidt:2019} or approximating the unknown energy functional and its derivatives \cite{Snyder:2012, Snyder:2013, Li:2016a, Li:2016b, Yao:2016, Seino:2018, Nelson:2019, Golub:2019, Nudejima:2019}. More recent works bypass the Kohn-Sham-solution scheme by directly learning the mapping between material parameters and ground-state properties \cite{Hansen:2013, Schutt:2014, Hansen:2015, Schutt:2017, Brockherde:2017, Bogojeski:2018,  Ryczko:2018, Schmidt:2018, Pilati:2019, Custodio:2019, Ryczko:2019, Zepeda:2019}, or constructing the ground-state wavefunction from a corresponding density distribution~\cite{Moreno:2019}. Despite these advancements, previous approaches are either limited by complicated, non-scalable networks, suffer from inefficient training data generation or struggle in applications to the different physical phases of the used models.

In this work, we take an approach that follows three guiding principles: 
(i) implicit knowledge representation is the key strength of neural networks. Therefore, we use a \emph{neural network to implicitely represent the (minimized) DFT functional}. 
(ii) We aim at solving for phases of quantum matter beyond the band structure paradigm. To that end, we train the neural network to directly \emph{output correlation functions}~\cite{Moreno:2019}, which can be used to characterize phases and phase transitions. 
(iii) A balanced dataset is the key challenge as data acquisition -- theoretical or experimental -- is costly. In such settings, \emph{active learning schemes}~\cite{Settles:2009, Gubaev:2019, Sivaraman:2019, Yao:2020, Teichert:2020} offer better results with fewer training instances. In general, active learning describes an algorithm which can actively choose the data it wants to learn from during training. Here we employ a procedure inspired by active learning, to incorporate data from different system sizes in the a priori data generation. In particular, our approach uses costly data of larger systems only in situations where large finite-size effects are detected.

\begin{figure*}
    \includegraphics[width=\linewidth]{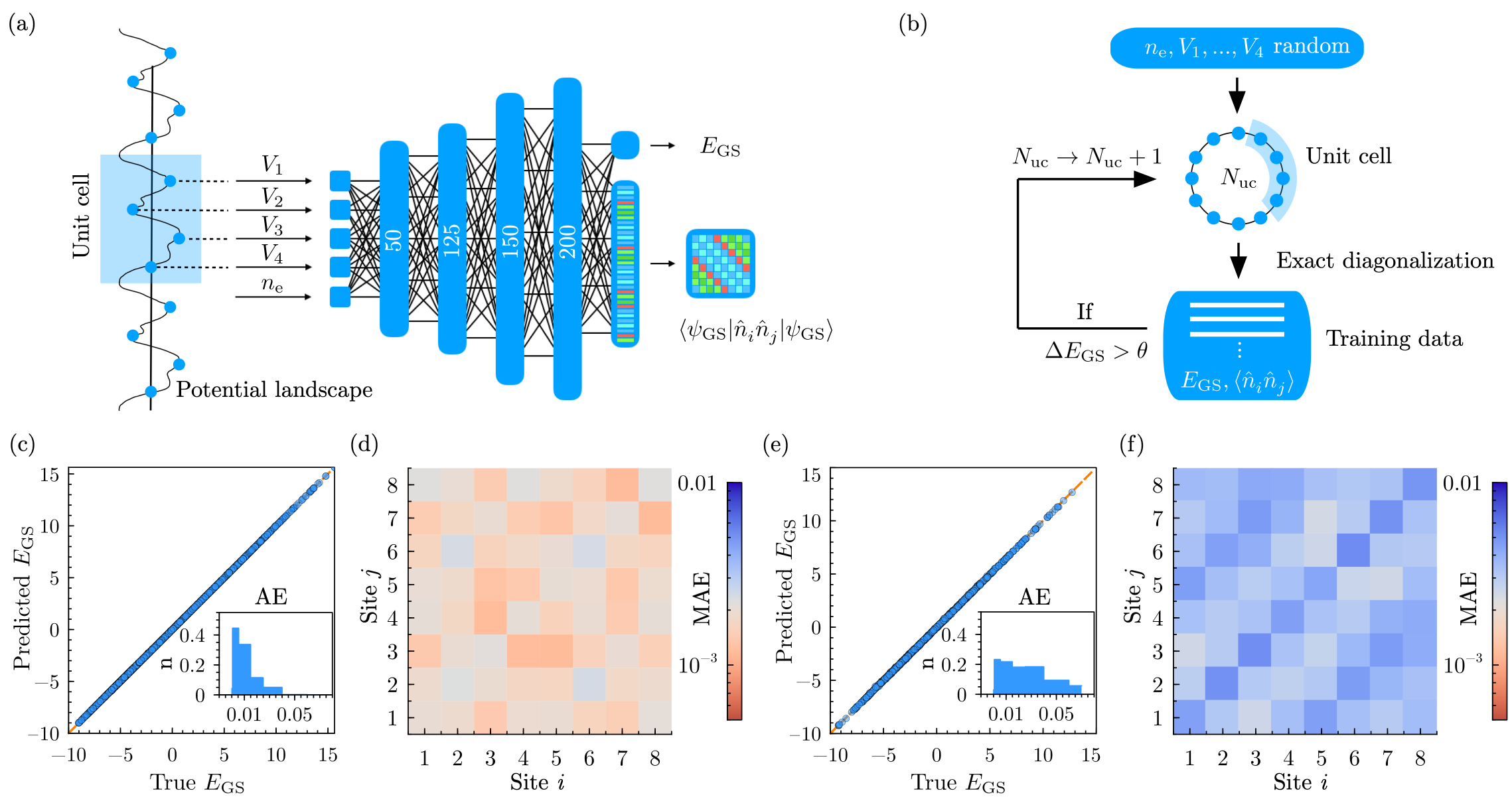}
    \caption{\label{fig:setup} (a) Schematic of the dense neural network used to learn the map from unit-cell potentials and filling to ground-state energy and density-density correlators. (b) Data generation inspired by active learning allows to check the energy deviations of different system sizes and continuing to larger systems only if necessary. (c) Exact versus predicted ground-state energies on the test data set of the actively learning network (ALN). The inset shows the absolute error on the test-energy values as a density histogram $n$. (d) Mean absolute error per correlator entry on the test data set for the ALN. (e) Exact versus predicted ground-state energies for the network trained on a single system size (PLN), with the inset showing the density $n$ of absolute errors on the test data set. (f) Mean absolute error per correlator entry on the test data set for the PLN.}
\end{figure*}

Figures~\ref{fig:setup}~(a) and~(b) summarize our model, neural network, and workflow. We choose to work with a one-dimensional lattice model of spinless fermions. The hopping and interaction terms of the model define our `universal Schr{\"o}dinger equation' and are therefore left unaltered throughout the study. Input to the neural network is the problem-specific potential and particle number. Its output is the ground-state energy $E_{\rm GS}$ as well as the density-density correlation function. We start by introducing the employed learning scheme and demonstrate the quantitative accuracy of network predictions, after training it on random potentials.
The active training shows superior performance compared to conventional training. We obtain mean squared errors of the energy of $3.08 \cdot 10^{-4} $ in units of the hopping integral. Finally, we apply the trained network to a topological and a symmetry-breaking phase transition. Our results demonstrate a scalable architecture, able to capture interacting lattice models, with successful applications to structured phases.

\section{Model}
While density functional theory was originally formulated as a continuum theory, it has also been successfully applied to lattice models~\cite{Schonhammer:1995}. 
We consider a Hamiltonian for spinless fermions on a one-dimensional lattice with sites labelled by $i=1,\cdots, L$ under periodic boundary conditions,
\begin{align}
    \hat{H} =& -t \sum_i \left( \hat{c}_i^{\dagger} \hat{c}^{}_{i+1} + \text{h.c.} \right) + U \sum_i \hat{n}^{}_i \hat{n}^{}_{i+1} \notag \\
    &+ U' \sum_i \hat{n}^{}_i \hat{n}^{}_{i+2} + \sum_i V^{}_i \hat{n}^{}_i,
    \label{eq:Ham}
\end{align}
where $\hat{c}_i^{\dagger}$ and $\hat{c}^{}_{i}$ are the fermion creation and annihilation operators on site $i$ and $\hat{n}^{}_i = \hat{c}_i^{\dagger} \hat{c}^{}_{i}$ is the corresponding density operator. Nearest-neighbor hopping is parametrized by $t$, which will serve as the energy unit throughout. The particles are subject to a repulsive interaction on nearest- and next-nearest-neighbor sites which we fix to $U=1$ and $U'=0.5$ so as to model a lattice analogue of the Coulomb interaction. This parameter choice places the system in a metallic, but strongly correlated phase in absence of a potential $V^{}_i$, even at half filling~\cite{Markhof:2018}. 

Motivated by the Hohenberg-Kohn theorems, we consider the kinetic term and the electron-electron interactions as universal, such that the external or ionic potential $\hat{V}_{\text{ext}}=\sum_i V_i \hat{n}_i$ together with the filling uniquely determine the ground state and all of its properties. We only consider potentials with periodicity of four sites. That is, the four values $V^{}_{i}$, $i = 1, ..., 4$, completely specify the Hamiltonian for any lattice size $L=4N_{\mathrm{uc}}$, with $N_{\mathrm{uc}}$ the number of unit cells and $V^{}_{i}=V^{}_{i+4}$ for all $i$. This four-site unit cell can be thought of as the discretized unit cell of a periodic crystal, while $V^{}_{i}$ is the ionic potential in this analogy. We restrict it to the range $V_i \in \left[-4,4\right]$.
We further denote by the real number $0<n_{\mathrm{e}}<4$ the particle filling per unit cell.
We emphasize that despite imposing this periodicity to the potential, our approach is able to capture phases which spontaneously break the four-site translation symmetry.

\section{Learning}
The supervised-machine-learning algorithm we use bypasses the Kohn-Sham scheme by directly learning the map from the external parameters $n^{}_{\mathrm{e}}$ and $V^{}_{i}$ to the corresponding ground-state energy and density-density correlators $\langle \hat{n}^{}_{i}\hat{n}^{}_{j} \rangle_{\mathrm{GS}}$. The density-density correlators are calculated for two adjacent unit cells~\footnote{In order to also use small systems ($N_{\mathrm{uc}}=4$) with periodic boundary conditions as part of the training data, at most two adjacent unit cells from the interior of the density-density correlator are a valid representative.}. The chosen fully connected neural network~\cite{Chollet:2015} consists of four hidden layers which increase in size towards the output as depicted in Fig.~\ref{fig:setup} (a) (see also Appendix~\ref{app:nn_parameters}).

A central challenge in machine learning is unbiased and efficient data generation; one usually deals with limited computational or experimental resources. Here, we generate data by finite-size exact diagonalization (ED) of systems with randomly chosen $n^{}_{\mathrm{e}},\ V_i$. In order to reduce finite size effects, these examples should naively be generated with as large systems as possible. However, the computational cost for data generation with ED grows exponentially with system size. For this reason, we employ a procedure inspired by \emph{active learning}, performing costly large system ED, as depicted in Fig. \ref{fig:setup} (b), only if necessary. Using random values for $n_{\mathrm{e}}$ and  $V_i$ and starting with a comparison between $N_{\mathrm{uc}}=3,4$, the scheme iteratively computes larger systems until the finite size deviation between ground-state energies lies below a priorly chosen threshold $\theta$. Correspondingly, the fast computation of smaller systems is used as often as possible, while providing more accurate data in critical cases (see Appendix~\ref{app:training_data}). The samples are further augmented by applying translations within the unit cell and inversion, allowing the network to capture the symmetries of the universal part of the Hamiltonian.

We contrast the active learning approach outlined above with a passive learning scheme using training data generated for systems of fixed size $N_{\mathrm{uc}}=5$, with filling $n^{}_{\mathrm{e}}$ and on-site potentials $V^{}_{i}$ chosen randomly. This system size is still solvable efficiently by ED, while sufficiently reducing finite size effects. The data are again symmetry-augmented. Both learning procedures were run with an equal time budget to ensure comparability.

\begin{figure}
    \includegraphics[width=\linewidth]{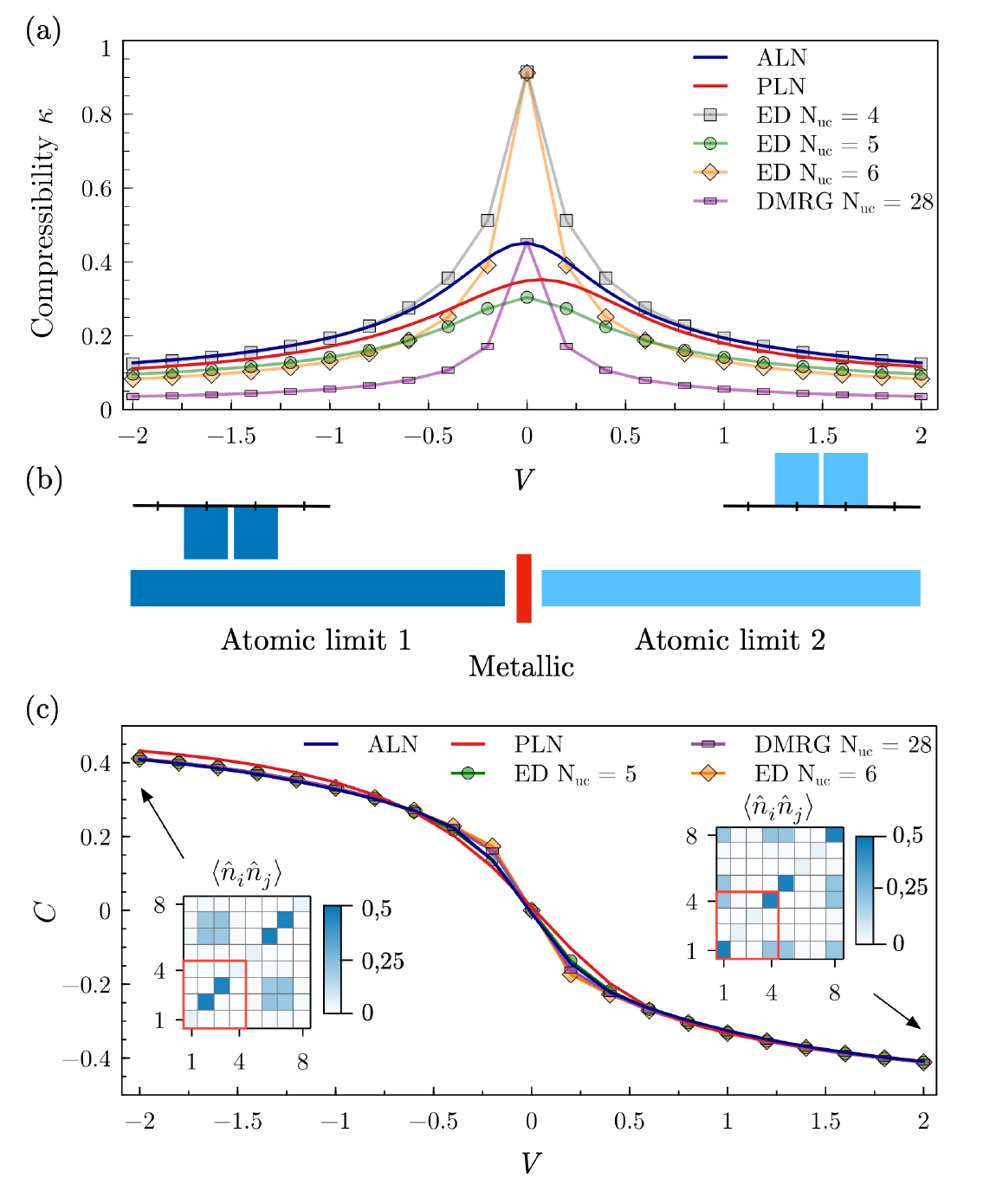}
    \caption{\label{fig:SSH}
    Neural network results for a transition between different (obstructed) atomic limit insulators. (a) Compressibility $\kappa$ for various potential strengths at quarter filling, as calculated from the actively and passively learned neural network, exact diagonalization (ED) and density matrix renormalization group (DMRG) of several system sizes. (b) Schematic depiction of the potential in the four-site unit cell: depending on the strength and sign, two obstructed atomic limits and a metallic phase can be realized. (c) Corresponding observable $C$ as calculated from the 8x8 density-density correlator for the same numerical methods as used for the compressibility. The insets show the correlator as obtained from the ALN in the first (lower left) and second atomic limit (upper right) with the unit cell depicted in red.}
\end{figure}

A mean absolute error loss function is then optimized to obtain the weights and biases of the actively (ALN) and passively (PLN) learning neural network. The resulting performance is evaluated on unseen data, consisting of 20 \% of the full data set. Overfitting was avoided for both systems by suitable hyperparameter choices~(see Appendix~\ref{app:training_performance}). The absence of significant deviations in the energy correlation plot in Fig.~\ref{fig:setup}~(c) shows that the ALN performs well on random data, with an absolute error peaked at $1.2\cdot 10^{-2}$. Similarly, Fig.~\ref{fig:setup}~(d) shows only small errors in the correlator prediction, with an overall mean absolute error of $1.8\cdot10^{-3}$. The PLN performs worse in predicting energy values and correlators, as shown in the inset of Fig.~\ref{fig:setup}~(e) and in Fig.~\ref{fig:setup}~(f), with errors at least twice as large as in the case of the ALN. This comparison shows the advantage of intelligent data generation at fixed computational time budget.  

Random potentials rarely represent a relevant physical scenario. We therefore present in the following how the randomly trained ALN outperforms the PLN for structured systems, where $V_i$ obey further symmetries. Interestingly, including the smaller samples, which are suffering from finite-size effects, with a reduced sample weight in the ALN training leads to a more stable model prediction, in particular for physical systems (see Appendix~\ref{app:training_data}). We attribute this feature to a regularization of the model.

\section{Learnability of obstructed atomic limits}
We consider the  model introduced in Eq.~\eqref{eq:Ham} for a potential choice $(V_1,V_2,V_3,V_4)=(0,V,V,0)$ at quarter filling ($n_{\mathrm{e}} = 1$).
The system is metallic for $V=0$ which separates two distinct insulating phases for  $V> 0$ and $V<0$.
 For $V>0$ ($V<0$), Wannier functions are localized between the unit cells (in the middle of the unit cell). This corresponds to two topologically distinct (obstructed) atomic-limit insulators, with the intra-unit-cell hopping being effectively reduced (enhanced) compared to the inter-unit-cell hopping. The insulating nature of these phases can be shown by calculating the compressibility
\begin{equation}
    \kappa = \frac{1}{n^{2}_{\mathrm{e}}} \left(\frac{\partial^2 E_{\mathrm{GS}}(n^{}_{\mathrm{e}})}{\partial n^{2}_{\mathrm{e}}}\right)^{-1},
    \label{eq:compressibility}
\end{equation}
where $n^{}_{\mathrm{e}}$ is the electron filling and $E_{\mathrm{GS}}(n^{}_{\mathrm{e}})$ the corresponding ground-state energy. Figure~\ref{fig:SSH}~(a) shows that, as one approaches the critical metallic state around $V = 0$, $\kappa$ increases rapidly. 
We emphasize that since $\kappa$ is the second derivative of the energy, it is extremely susceptible to errors. Note, further, that the ED data show a strong even-odd effect in $N_{\mathrm{uc}}$.
We also calculated $\kappa(V)$ with the density matrix renormalization group algorithm (DMRG)~\footnote{Calculations were performed using the TeNPy Library (version 0.5.0), for a finite but periodic lattice with maximum bond dimension $\chi_{\mathrm{max}} = 2000$, max. truncation error $\epsilon = 10\mathrm{e}{-8}$ and energy convergence criterion $\Delta E = 10\mathrm{e}{-6}$.}\cite{Hauschild:2018} for $N_{\mathrm{uc}}=28$.
Compared with these exact results, the ALN produces a meaningful $\kappa(V)$ with a peak value $\kappa(V=0)$ interpolating between the results of even and odd $N_{\mathrm{uc}}$. On the contrary, the PLN is worse with a less pronounced and non-symmetric peak. Even though only trained with at most $N_{\mathrm{uc}} = 6$ data, we see that the ALN also resembles the $N_{\mathrm{uc}} = 28$ DMRG result reasonably well.

\begin{figure}
    \includegraphics[width=\linewidth]{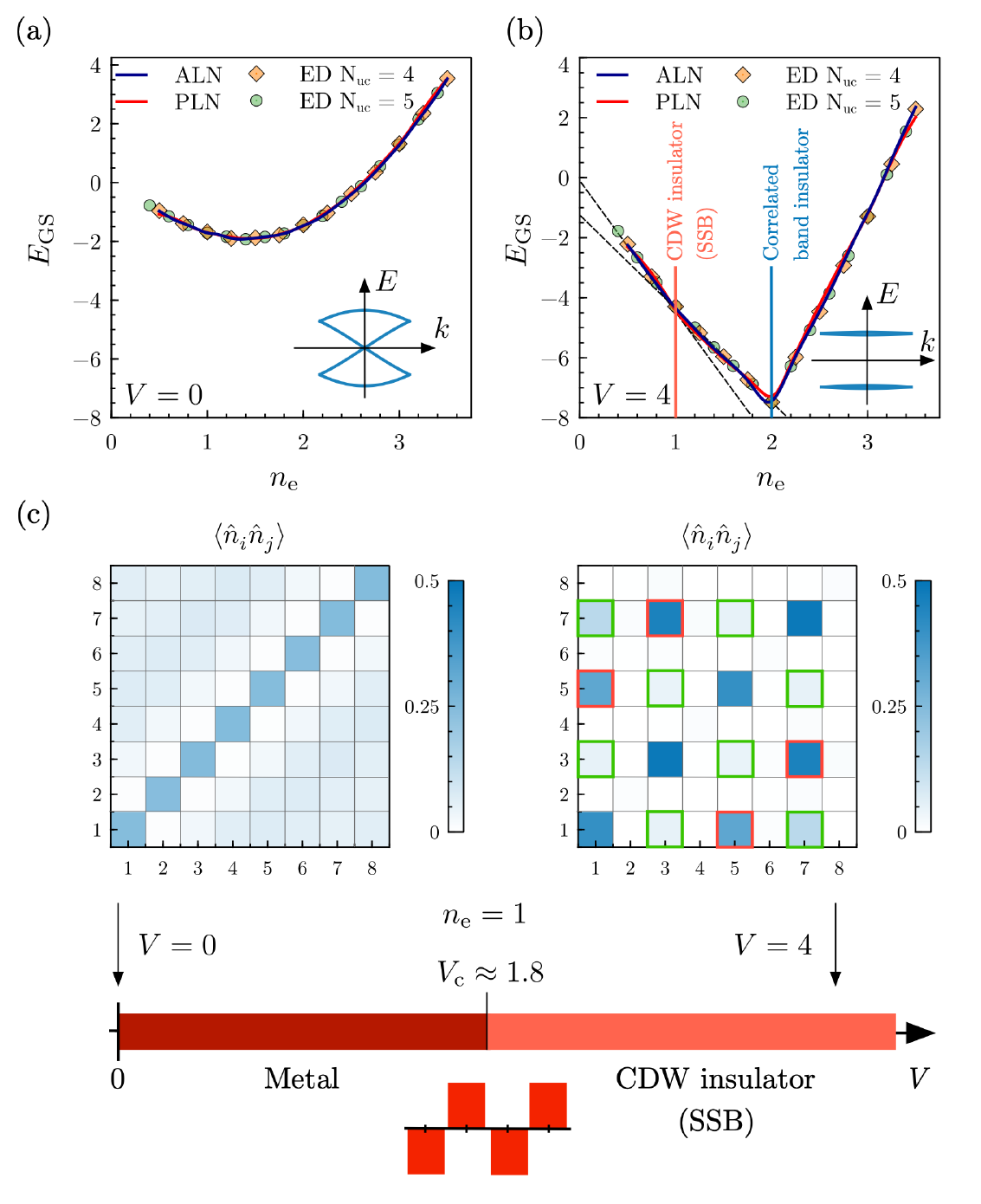}
    \caption{\label{fig:SSB} 
    Neural network results for a spontaneous symmetry-breaking phase.
    (a) Ground-state energy for $V = 0$ for several electron fillings $n^{}_{\mathrm{e}}$ as calculated from the actively and passively learned neural network and ED. The inset displays the noninteracting band structure. (b) Ground-state energy as a function of the electron filling $n^{}_{\mathrm{e}}$ in the symmetry broken phase ($V = 4$). The kink at $n_{\mathrm{e}} = 1$ signals an interaction-induced incompressible phase. The inset reveals the band flattening of the non-interacting system. (c) Phase diagram, schematic of the potential, and density-density correlation functions. The latter are obtained from the ALN for $V = 0$ (left) and $V = 4$ (right). Off-diagonal terms, whose inequivalence signals the symmetry-breaking phase (right) are highlighted in red and green.}
\end{figure}

The location of Wannier centers can be used to differentiate between the two phases. Defining $C = \left(\langle \hat{n}_2 \hat{n}_2 \rangle - \langle \hat{n}_2 \hat{n}_3 \rangle\right) - \left(\langle \hat{n}_4 \hat{n}_4 \rangle - \langle \hat{n}_4 \hat{n}_5 \rangle\right)$~\footnote{Note that we choose $C$ to highlight that two neighboring sites are always occupied by just one electron. That distinguishes it from the band insulator case of filling $n_{\mathrm{e}} = 2$ and a spontaneously translation symmetry broken phase at $n_{\mathrm{e}} = 1$ (both $C = 0$).}, the trivial atomic limit with localization in the unit cell is obtained for $C > 0$, the phase transition happens at $C = 0$ and the non-trivial atomic limit has $C < 0$. Figure~\ref{fig:SSH}~(c) highlights that both networks are able to capture $C$ across the transition well, but the ALN results are markedly more accurate than the PLN results when compared with the ED and DMRG data. This supports the statement that our active learning scheme delivers quantitatively better results.

\section{Learnability of spontaneously symmetry-broken phases}
Spontaneous breaking of translation symmetry can be triggered by introducing a potential of the form $(V_1,V_2,V_3,V_4)=(-V,V,-V,V)$ at quarter filling ($n_{\mathrm{e}} = 1$). The symmetry-broken phase arises from the competition between the next-nearest-neighbor interaction $U'$ and the increasing potential $V$. The four-site translational symmetry of Hamiltonian~(\ref{eq:Ham}) is broken spontaneously at $V_c \approx 1.8$~(see Appendix~\ref{app:SSB_Vcrit}) by the two-site relations of the emerging degenerate ground-states. The metallic system shows a smooth dependence of $E_{\rm GS}$ on $n_{\mathrm{e}}$ around quarter filling [Fig.~\ref{fig:SSB}~(a)]. With increasing $V$, $E_{\rm GS}(n_{\mathrm{e}})$ develops a kink at $n_{\mathrm{e}} = 1$, signalling the emergence of the symmetry-broken charge-density wave (CDW) insulator [Fig.~\ref{fig:SSB}~(b)]. Both neural networks represent the different phases very well, the deviations at small fillings are attributed to the limited amount of training samples in this limit.

Figure~\ref{fig:SSB}~(c) shows that the correlation in the metallic phase is short-ranged and fast decaying, whereas the symmetry broken phase possesses a distinct order. The corresponding order parameter $\langle C_{\mathrm{SSB}} \rangle= \frac{1}{N_{\mathrm{uc}}} \langle\sum_{i} (-1)^{i} \hat{n}^{}_{2i+1} \rangle$ is, however, zero, since the two degenerate ground states have opposite imbalance in electron density between the first and third site of each unit cell. Instead, the order can be diagnosed by computing the square of the order parameter from the density-density correlation functions. This amounts to $\langle C_{\mathrm{SSB}}
^2\rangle= \frac{2}{N_{\mathrm{uc}}^2} \sum_{i \neq j}\langle\hat{n}^{}_{4i+1}\hat{n}^{}_{4j+1}\rangle - \frac{2}{N_{\mathrm{uc}}^2} \sum_{i,j}\langle\hat{n}^{}_{4i+1}\hat{n}^{}_{4j+3}\rangle + C_0$, with an overall shift $C_0 = \frac{1}{N_{\mathrm{uc}}^2} \langle\sum_{i}  \hat{n}^{}_{2i+1} \rangle$. Its nonzero value in the symmetry broken phase is implied by the inequivalence between the first (+) and second (-) term in the above expression, highlighted in the density-density correlator in Fig.~\ref{fig:SSB}~(c) by the red (+) and green (-) squares. This behavior is well captured by the ALN, producing quantitatively accurate correlations in both phases.

\section{Conclusion}
We presented a supervised learning approach for lattice DFT, bypassing the Kohn-Sham solution scheme. Employing a procedure inspired by active learning allowed us to improve our results at fixed computational cost regarding data generation. Focussing on correlation functions on a subsystem and taking only the potential landscape in the unit cell and particle number as input results in a scalable architecture. Besides verification of our algorithm on unseen random potentials, we demonstrated that the trained networks reliably solve for different structured phases. 

Looking ahead, it is highly desirable to construct similar implicit (neural network) representations of DFT for systems in continuous space and higher dimensions, in particular to attack the electronic structure problem in strongly correlated regimes. The main challenge is the generation of valid and balanced data sets, and the incorporation of data from various sources, including conventional DFT, Monte Carlo calculations, experiments, and future quantum simulation devices. Two concepts on which our study relies, (1) focus on correlation functions instead of quantum states and (2) the use of efficient learning, should prove useful in this future venture. 
\newline

\acknowledgements
We thank Giuseppe Carleo and Xi Dai for insightful discussions. 
This project has received funding from the European Research Council (ERC) under the European Union’s Horizon 2020 research and innovation program (ERC-StG-Neupert-757867-PARATOP).

\appendix
\section{Network parameters}
\label{app:nn_parameters}

The supervised-machine-learning algorithm we propose in this paper uses dense neural networks of identical architecture, for both the active and passive learning scheme. The layers are connected by Softplus$(x)= \ln{\left(1+e^x\right)}$ activation functions, except for the output layer. The last layer takes into account that correlator values and ground-state energies have different ranges, by employing a linear activation function. 
Table \ref{tab:nn_params} indicates the relevant parameters used in this paper.

\begin{table}[h]
    \caption{\label{tab:nn_params} Relevant parameters used to create the neural networks in this paper.}
    
    \begin{tabular}{|c|c|}\hline
    Parameter & Value \\\hline
    Neurons Layer 1 & 50 \\
    Activation 1 & Softplus \\
    Weight init. 1 & lecun\_uniform \\
    Neurons Layer 2 & 125 \\
    Activation 2 & Softplus \\
    Weight init. 2 & lecun\_uniform \\
    Neurons Layer 3 & 150 \\
    Activation 3 & Softplus \\
    Weight init. 3 & lecun\_uniform \\
    Neurons Layer 4 & 200 \\
    Activation 4 & Softplus \\
    Weight init. 4 & lecun\_uniform \\
    Neurons Layer 5 & 65 \\
    Activation 5 & Linear \\
    Weight init. 5 & lecun\_uniform \\\hline

    Optimizer & Adam \\
    Batch size & 100 \\
    Learning rate & 0.001\\
    Epochs & 1500 \\\hline
    \end{tabular}
\end{table}

\section{\label{app:training_data}Training Data}

The choice of training examples is crucial in a machine learning setting, as data is precious. An intelligent data generation procedure is advantageous if computational time is finite, removing the necessity to always calculate as large systems as possible. Naively, the latter is the way to go in order to avoid finite size biases in the training examples. We contrast these two approaches as active and passive learning schemes.

The naive approach generates data with exact diagonalization of a five unit cell system, with electron fillings $n^{}_{\mathrm{e}}$ and potentials $V^{}_{i}$ in the unit cell chosen randomly. The potentials are in the range $\vert V^{}_{i} \vert \leq 4$, and the filling can assume values between $0.6 \leq n^{}_{\mathrm{e}} \leq 3.4$ electrons per unit cell. Data augmentation is then used to enlarge the dataset and allow the neural network to capture the underlying symmetries of the physical problem. This means applying translations $V^{}_{i} \rightarrow V^{}_{i+1}$ and inversion within the unit cell, where the former leads to a shift of the correlator data. The resulting dataset consists of 12.020 pairs of fillings and external potentials, mapped to the corresponding ground-state energies and density-density correlators. The split in training, validation and test sets is illustrated in Tab. \ref{tab:data_pln}.

\begin{table}[h]
    \caption{\label{tab:data_pln} Data for the passively trained neural network.}
    \begin{tabular}{|c|c|}\hline
    Parameter & Value \\\hline
    Total number of samples & 12020 \\
    Number of training samples & 7212 \\
    Number of validation samples & 2404 \\
    Number of test samples & 2404 \\\hline
    
    Number of samples from 20 site ED & 12020 \\
    Sample weight & 1 \\\hline
    \end{tabular}
\end{table}

However, large system diagonalization is not always necessary to obtain accurate data. This is the basis for an active learning scheme, which performs the diagonalization of large systems only if strong finite size effects are detected. The input to this procedure is the random choice of electron fillings $n^{}_{\mathrm{e}}$ and potentials $V^{}_{i}$ in the unit cell. The former assuming values between $0.5 \leq n^{}_{\mathrm{e}} \leq 3.5$ electrons per unit cell and the latter being chosen in the range $\vert V^{}_{i} \vert \leq 4$. The training data for both networks was chosen as $\vert V^{}_{i} \vert \leq 4$ in order to ensure that the transition to the symmetry broken phase lies within the trained potential range.

If the ground-state energy for a $N_{\mathrm{uc}} = 3$ and $N_{\mathrm{uc}} = 4$ unit cell system, calculated with exact diagonalization, deviates more than a priorly chosen threshold $\theta$, a system with one additional unit cell is being calculated. This procedure is repeated up to $N_{\mathrm{uc}} = 6$ unit cells if necessary. This means that the neural network can query a larger system whenever the deviation of ground-state energies exceeds the chosen threshold $\theta$, thereby reducing finite size effects. Naively one would remove the small inaccurate samples from the dataset, however we found that including them with a reduced sample weight yields similar results on random test data. However, we obtain better and more stable results on physical phases, meaning that the presence of the less accurate data points acts as a regularization of the model. As we strive for meaningful results on physical phases and not on random potentials, we consider this approach to be more promising. Data augmentation is again used not only to enlarge the dataset to 74.500 input-output data pairs (see Tab. \ref{tab:data_aln}), but also to allow the network to capture the underlying symmetries. 

\begin{table}[h]
    \centering
    \caption{\label{tab:data_aln} Data for the actively trained neural network.}
    \begin{tabular}{|c|c|}\hline
    Parameter & Value \\\hline
    Total number of samples & 74500 \\
    Number of training samples & 44700 \\
    Number of validation samples & 14900 \\
    Number of test samples & 14900 \\\hline
    
    Number of samples from 16 site ED & 68850 \\
    Number of samples from 20 site ED & 5500 \\
    Number of samples from 24 site ED & 150 \\
    Sample weight 16 site data & 1\footnote{This weight is increased to 3 if no 20 or 24 site system had to be calculated for this sample.} \\
    Sample weight 20 site data & 2\footnote{This weight is increased to 3 if no 24 site system had to be calculated for this sample.}\\
    Sample weight 24 site data & 3\\\hline
    \end{tabular}
\end{table}

The choice of the threshold $\theta$ is a crucial parameter of the active learning scheme. We therefore investigate the implications of different thresholds (see Tab. \ref{tab:theta_datasets}), and test the performance of the trained models on unseen examples. These examples are not part of the training set and were generated with random fillings $n^{}_{\mathrm{e}}$ and potentials $V^{}_{i}$ in the range $\vert V^{}_{i} \vert \leq 4$ for systems of $N_{\mathrm{uc}} = 5, 6$.

\squeezetable
\begingroup
\begin{table}[h]
    \centering
    \caption{\label{tab:theta_datasets} Number of samples for different choices of the threshold $\theta$, calculated with exact diagonalization and an equal time budget for each parameter choice.}
    \begin{tabular}{|c|c|c|c|c|c|c|}\hline
    $N_{\mathrm{uc}}$ & $\theta = 0.0$ & $\theta = 0.001$& $\theta = 0.0025$& $\theta = 0.005$ & $\theta = 0.01$ & $\theta = 1.5$\\\hline
    4 & 1350 & 4200 & 13500 & 18850 & 81075 & 83250  \\
    5 & 1335 & 575 & 1030 & 790 & 850 & 0  \\
    6 & 245 & 10 & 30 & 15 & 20 & 0  \\\hline
    \end{tabular}
\end{table}
\endgroup

\begin{figure}[h]
    \includegraphics[width=\linewidth]{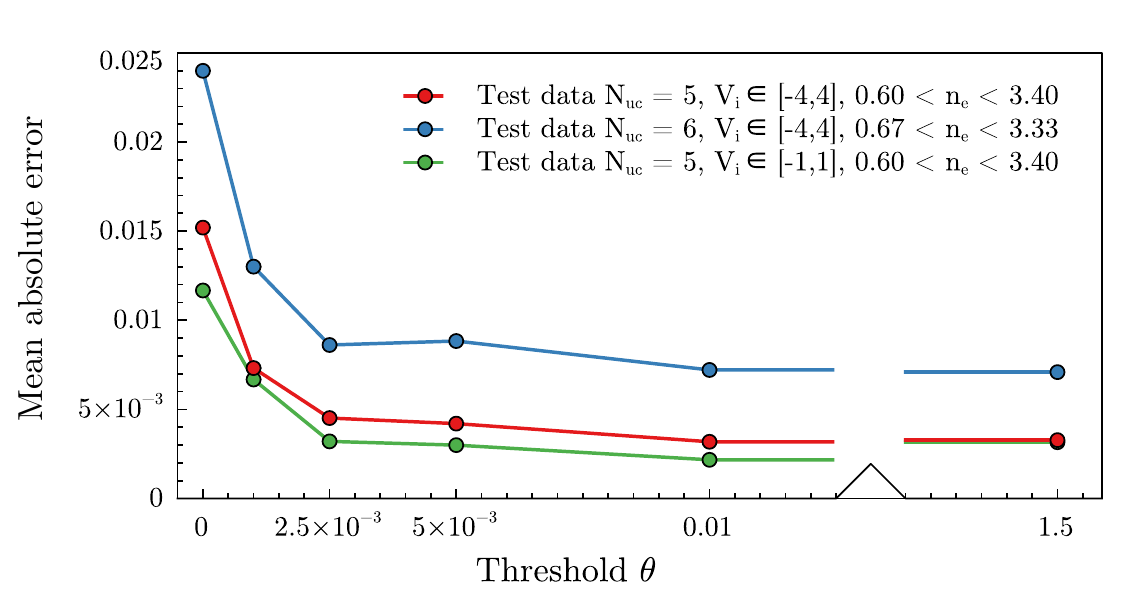}
    \caption{\label{fig:theta_plot}Mean absolute error of neural network predictions for various datasets, trained on examples generated with different threshold $\theta$ as illustrated in Tab \ref{tab:theta_datasets}. Random potentials were used to generate the evaluation data, with a $N_{\mathrm{uc}} = 5, \vert V^{}_{i} \vert \leq 4$ dataset consisting of 5600 samples, $N_{\mathrm{uc}} = 6, \vert V^{}_{i} \vert \leq 4$ containing 220 datapoints and 1205 $N_{\mathrm{uc}} = 5, \vert V^{}_{i} \vert \leq 1$ input - output pairs. The evaluation sets were not used in the training process of the neural networks.}
\end{figure}

The resulting mean absolute error is presented in Fig. \ref{fig:theta_plot}, indicating that the error decreases with increasing $\theta$. When considering random potentials, it is therefore advantageous to use training data generated with as large $\theta$ as possible, resulting in the largest possible dataset. Consequently, this means larger systems with $N_{\mathrm{uc}} = 5, 6$ are never calculated. 

The random data generation with potentials in the range $\vert V^{}_{i} \vert \leq 4$ causes however mostly situations where the electrons are localized in the potential landscape, due to large potentials $V^{}_{i}$. Finite size deviations are however negligible in such a case. This means that a good performance on this data does not necessarily correspond to a good elimination of finite size effects in a realistic physical potential. Figure \ref{fig:theta_plot} highlights this with the evaluation of a dataset with potentials $\vert V^{}_{i} \vert \leq 1$, with the minimum of the obtained mean absolute error shifting towards smaller values of $\theta$.

Consequently, a trade-off is needed between the total number of training samples and the number of samples with $N_{\mathrm{uc}} = 5, 6$. The best compromise is reached for $\theta = 0.0025$, with a large enough size of the dataset, while containing significant information about larger systems. This parameter was therefore used to generate the final dataset used to train the actively learning neural network in this paper.

\section{Training Performance}
\label{app:training_performance}

The actively and passively constructed datasets are split into training, validation and test sets. Achieving consistent performance on the optimized and unseen data indicates that the network has not been overfitted. The corresponding results for both training schemes are presented in Tab. \ref{tab:per_pln} and \ref{tab:per_aln}, highlighting that the intended mapping from electron fillings and potentials to ground-state energy and density-density correlator is well captured.

\begin{table}[h]
    \caption{\label{tab:per_pln}Performance of the passively trained neural network. The last table section shows the performance on the ALN test dataset.}
    \centering
    \begin{tabular}{|c|c|}\hline
    Parameter & Value \\\hline
    MAE on $E_{\mathrm{GS}}^{\mathrm{test}}$ & 2.8e-2 \\
    MSE on $E_{\mathrm{GS}}^{\mathrm{test}}$ & 1.5e-3 \\
    MAE on $\langle \hat{n}^{}_{i}\hat{n}^{}_{j} \rangle_{\mathrm{GS}}^{\mathrm{test}}$ & 3.6e-3 \\
    MSE on $\langle \hat{n}^{}_{i}\hat{n}^{}_{j} \rangle_{\mathrm{GS}}^{\mathrm{test}}$ & 5.9e-5 \\\hline
    MAE on $E_{\mathrm{GS}}^{\mathrm{validation}}$ & 2.6e-2 \\
    MSE on $E_{\mathrm{GS}}^{\mathrm{validation}}$ & 1.3e-3 \\
    MAE on $\langle \hat{n}^{}_{i}\hat{n}^{}_{j} \rangle_{\mathrm{GS}}^{\mathrm{validation}}$ & 3.5e-3 \\
    MSE on $\langle \hat{n}^{}_{i}\hat{n}^{}_{j} \rangle_{\mathrm{GS}}^{\mathrm{validation}}$ & 4.9e-5 \\\hline
    MAE on $E_{\mathrm{GS}}^{\mathrm{training}}$ & 2.5e-3 \\
    MSE on $E_{\mathrm{GS}}^{\mathrm{training}}$ & 1.2e-3 \\
    MAE on $\langle \hat{n}^{}_{i}\hat{n}^{}_{j} \rangle_{\mathrm{GS}}^{\mathrm{training}}$ & 3.2e-3 \\
    MSE on $\langle \hat{n}^{}_{i}\hat{n}^{}_{j} \rangle_{\mathrm{GS}}^{\mathrm{training}}$ & 4.2e-5 \\\hline
    MAE on $E_{\mathrm{GS}}^{\mathrm{ALN test}}$ & 3.7e-2 \\
    MSE on $E_{\mathrm{GS}}^{\mathrm{ALN test}}$ & 2.7e-3 \\
    MAE on $\langle \hat{n}^{}_{i}\hat{n}^{}_{j} \rangle_{\mathrm{GS}}^{\mathrm{ALN test}}$ & 4.5e-3 \\
    MSE on $\langle \hat{n}^{}_{i}\hat{n}^{}_{j} \rangle_{\mathrm{GS}}^{\mathrm{ALN test}}$ & 7.0e-5 \\\hline
    \end{tabular}
\end{table}

\begin{table}[h]
    \caption{\label{tab:per_aln}Performance of the actively trained neural network. The last table section shows the performance on the PLN test dataset.}
    \centering
    \begin{tabular}{|c|c|}\hline
    Parameter & Value \\\hline
    MAE on $E_{\mathrm{GS}}^{\mathrm{test}}$ & 1.2e-2 \\
    MSE on $E_{\mathrm{GS}}^{\mathrm{test}}$ & 2.4e-4 \\
    MAE on $\langle \hat{n}^{}_{i}\hat{n}^{}_{j} \rangle_{\mathrm{GS}}^{\mathrm{test}}$ & 1.8e-3 \\
    MSE on $\langle \hat{n}^{}_{i}\hat{n}^{}_{j} \rangle_{\mathrm{GS}}^{\mathrm{test}}$ & 1.8e-5 \\\hline
    MAE on $E_{\mathrm{GS}}^{\mathrm{validation}}$ & 1.2e-2 \\
    MSE on $E_{\mathrm{GS}}^{\mathrm{validation}}$ & 2.3e-4 \\
    MAE on $\langle \hat{n}^{}_{i}\hat{n}^{}_{j} \rangle_{\mathrm{GS}}^{\mathrm{validation}}$ & 1.8e-3 \\
    MSE on $\langle \hat{n}^{}_{i}\hat{n}^{}_{j} \rangle_{\mathrm{GS}}^{\mathrm{validation}}$ & 1.5e-5 \\\hline
    MAE on $E_{\mathrm{GS}}^{\mathrm{training}}$ & 1.2e-2 \\
    MSE on $E_{\mathrm{GS}}^{\mathrm{training}}$ & 2.3e-4 \\
    MAE on $\langle \hat{n}^{}_{i}\hat{n}^{}_{j} \rangle_{\mathrm{GS}}^{\mathrm{training}}$ & 1.8e-3 \\
    MSE on $\langle \hat{n}^{}_{i}\hat{n}^{}_{j} \rangle_{\mathrm{GS}}^{\mathrm{training}}$ & 1.2e-5 \\\hline
    MAE on $E_{\mathrm{GS}}^{\mathrm{PLN test}}$ & 1.8e-2 \\
    MSE on $E_{\mathrm{GS}}^{\mathrm{PLN test}}$ & 5.2e-4 \\
    MAE on $\langle \hat{n}^{}_{i}\hat{n}^{}_{j} \rangle_{\mathrm{GS}}^{\mathrm{PLN test}}$ & 2.9e-3 \\
    MSE on $\langle \hat{n}^{}_{i}\hat{n}^{}_{j} \rangle_{\mathrm{GS}}^{\mathrm{PLN test}}$ & 3.6e-5 \\\hline
    \end{tabular}
\end{table}

\section{Determination of the point of spontaneous symmetry-breaking}
\label{app:SSB_Vcrit}

The four-site translational symmetry of the considered extended Hubbard model can be spontaneously broken by the introduction of an external potential $\hat{V}_{\text{ext}}=\sum_{i=1}^{4} V_i \hat{n}_i$. Choosing a potential of the form $V^{}_{1}=-V^{}_{2}=V^{}_{3}=-V^{}_{4} = -V$ causes a competition between next-nearest neighbor repulsion $U'$ and the potential $V$ at quarter filling. As a result, a symmetry-breaking phase with two degenerate ground-states emerges, corresponding to ordering the electrons either to site one or to site three in each unit cell.

We consider two approaches in order to identify the potential strength at which the spontaneous symmetry-breaking occurs. For each of these we investigate a finite size scaling plot, to derive the extrapolated point of the phase transition in the thermodynamic limit.

\begin{figure*}[ht]
    \includegraphics[width=\linewidth]{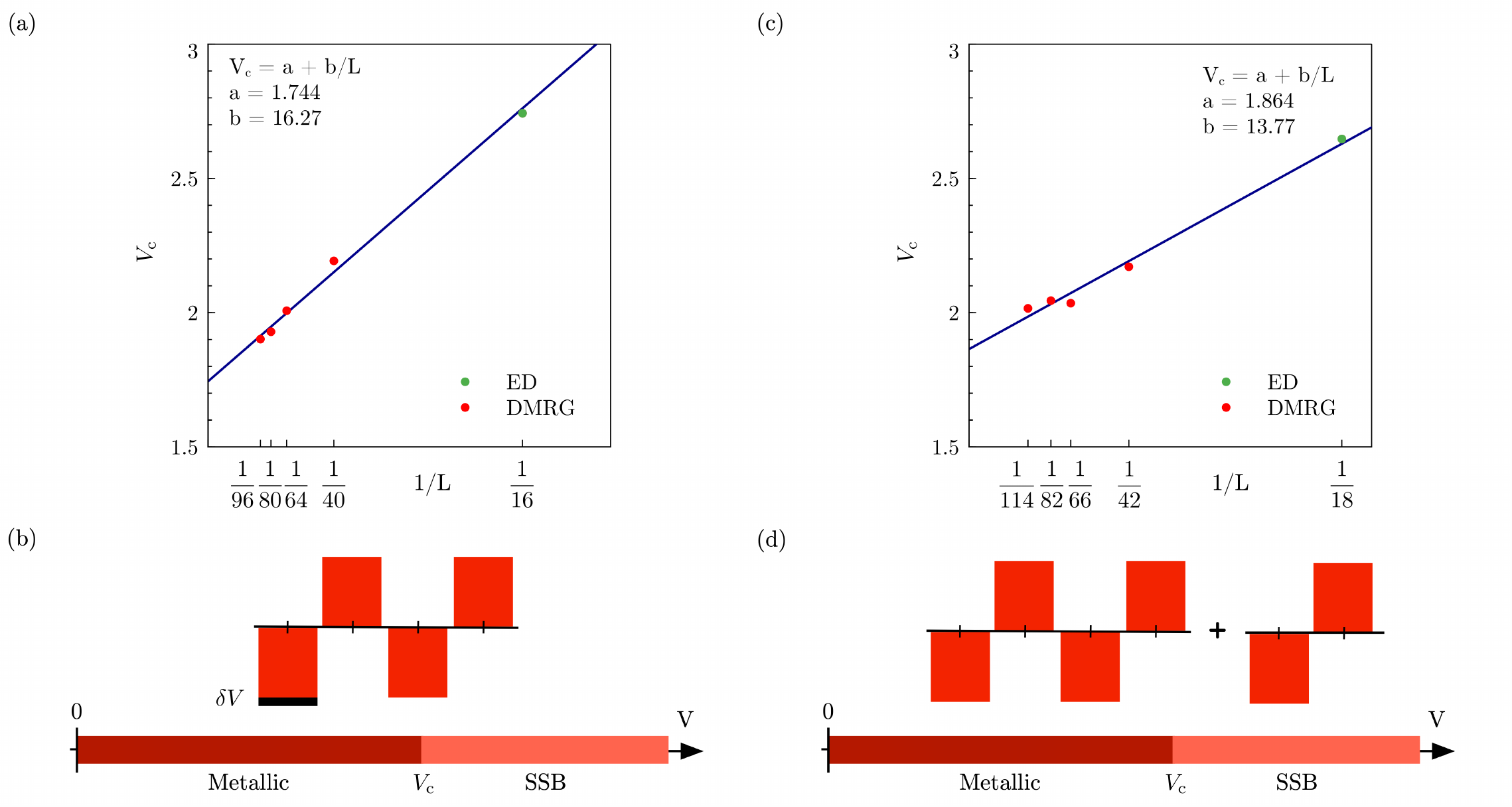}
    \caption{\label{fig:SSB_pt_point} (a) Finite size scaling plot of the critical potential $V_c$, obtained with a small potential $\delta V$ to break the ground-state degeneracy. Density matrix renormalization group (DMRG) calculations of open boundary conditions are combined with the exact diagonalization (ED) of a 16 site system. (b) Schematic of the potential landscape in the unit cell. (c) Finite size scaling plot of the critical potential $V_c$, obtained with an additional half unit cell and electron to break the ground-state degeneracy. Density matrix renormalization group (DMRG) calculations of open boundary conditions are combined with the exact diagonalization (ED) of an 18 site system. (d) Schematic of the potential landscape with additional two lattice sites.}
\end{figure*}

As stated above, the symmetry broken phase possesses a distinct order in the density-density correlator, occupying only sites with negative potential. By adding a small potential on one site, e.g. $V^{}_{1}=-V-\delta V,~ \delta V \ll V$, one of the degenerate ground-states is favoured. Consequently, the occupation $\langle \hat{n}^{}_{i} \rangle_{\mathrm{GS}}$ will concentrate on the first lattice site of each unit cell. We therefore investigate the difference in the occupation of site one and three, $C = \langle \hat{n}^{}_{1} \rangle_{\mathrm{GS}}-\langle \hat{n}^{}_{3} \rangle_{\mathrm{GS}}$. In the metallic phase for a vanishing potential $V = 0$, all sites are equally occupied, despite the small offset $\delta V$. This changes around the point of the phase transition, where explicitly one of the symmetry-breaking ground states is selected. The position of this jump in $C$ can then be studied with various system sizes and extrapolated to the thermodynamic limit. Since exact diagonalization already reaches computational boundaries for quite small systems, we additionally employ the density-matrix renormalization group algorithm (DMRG)~\footnote{Calculations were performed using the TeNPy Library (version 0.5.0), for a finite lattice with maximum bond dimension $\chi_{\mathrm{max}} = 1000$, max. truncation error $\epsilon = 10\mathrm{e}{-10}$ and energy convergence criterion $\Delta E = 10\mathrm{e}{-6}$.}\cite{Hauschild:2018}. In order to ensure convergence also for large systems, open boundary conditions are considered. The influence of the boundary can however be neglected when considering density-density correlations in the middle of the system. Figure \ref{fig:SSB_pt_point}(a) indicates the transition point as extrapolated from several system sizes, leading to a potential of $V_c = 1.744$.

Additionally, one can study the emergence of the spontaneously symmetry-breaking phase in the energy spectrum. The restriction to systems with open boundary conditions for DMRG calculations leads to a degeneracy of ground states of the order of the number of unit cells, compared to two states in the case of periodic boundaries. We therefore enlarge the system by two lattice sites and add one additional electron for the entire system, breaking the ground-state degeneracy. The transition to the symmetry broken phase can consequently be probed by calculating the gap $E_{\mathrm{GS}} - E_{\mathrm{1}}$ for several system sizes. The critical potential $V_c$ corresponds to the point of gap opening. The corresponding results are presented in Fig. \ref{fig:SSB_pt_point}(c), with an extrapolated transition point at $V_c = 1.864$. Taking the mean between both results gives a critical potential $V_c \approx 1.8$. Correspondingly, choosing the range of training potentials as $V_i \in \left[-4,4\right]$ secures that the neural networks can capture the phase transition.

\bibliography{references} 

\begin{thebibliography}{45}%
\makeatletter
\providecommand \@ifxundefined [1]{%
 \@ifx{#1\undefined}
}%
\providecommand \@ifnum [1]{%
 \ifnum #1\expandafter \@firstoftwo
 \else \expandafter \@secondoftwo
 \fi
}%
\providecommand \@ifx [1]{%
 \ifx #1\expandafter \@firstoftwo
 \else \expandafter \@secondoftwo
 \fi
}%
\providecommand \natexlab [1]{#1}%
\providecommand \enquote  [1]{``#1''}%
\providecommand \bibnamefont  [1]{#1}%
\providecommand \bibfnamefont [1]{#1}%
\providecommand \citenamefont [1]{#1}%
\providecommand \href@noop [0]{\@secondoftwo}%
\providecommand \href [0]{\begingroup \@sanitize@url \@href}%
\providecommand \@href[1]{\@@startlink{#1}\@@href}%
\providecommand \@@href[1]{\endgroup#1\@@endlink}%
\providecommand \@sanitize@url [0]{\catcode `\\12\catcode `\$12\catcode
  `\&12\catcode `\#12\catcode `\^12\catcode `\_12\catcode `\%12\relax}%
\providecommand \@@startlink[1]{}%
\providecommand \@@endlink[0]{}%
\providecommand \url  [0]{\begingroup\@sanitize@url \@url }%
\providecommand \@url [1]{\endgroup\@href {#1}{\urlprefix }}%
\providecommand \urlprefix  [0]{URL }%
\providecommand \Eprint [0]{\href }%
\providecommand \doibase [0]{http://dx.doi.org/}%
\providecommand \selectlanguage [0]{\@gobble}%
\providecommand \bibinfo  [0]{\@secondoftwo}%
\providecommand \bibfield  [0]{\@secondoftwo}%
\providecommand \translation [1]{[#1]}%
\providecommand \BibitemOpen [0]{}%
\providecommand \bibitemStop [0]{}%
\providecommand \bibitemNoStop [0]{.\EOS\space}%
\providecommand \EOS [0]{\spacefactor3000\relax}%
\providecommand \BibitemShut  [1]{\csname bibitem#1\endcsname}%
\let\auto@bib@innerbib\@empty
\bibitem [{\citenamefont {Hohenberg}\ and\ \citenamefont
  {Kohn}(1964)}]{Hohenberg:1964}%
  \BibitemOpen
  \bibfield  {author} {\bibinfo {author} {\bibfnamefont {P.}~\bibnamefont
  {Hohenberg}}\ and\ \bibinfo {author} {\bibfnamefont {W.}~\bibnamefont
  {Kohn}},\ }\href {\doibase 10.1103/PhysRev.136.B864} {\bibfield  {journal}
  {\bibinfo  {journal} {Physical Review}\ }\textbf {\bibinfo {volume} {136}},\
  \bibinfo {pages} {B864} (\bibinfo {year} {1964})}\BibitemShut {NoStop}%
\bibitem [{\citenamefont {Kohn}\ and\ \citenamefont {Sham}(1965)}]{Kohn:1965}%
  \BibitemOpen
  \bibfield  {author} {\bibinfo {author} {\bibfnamefont {W.}~\bibnamefont
  {Kohn}}\ and\ \bibinfo {author} {\bibfnamefont {L.~J.}\ \bibnamefont
  {Sham}},\ }\href {\doibase 10.1103/PhysRev.140.A1133} {\bibfield  {journal}
  {\bibinfo  {journal} {Physical Review}\ }\textbf {\bibinfo {volume} {140}},\
  \bibinfo {pages} {A1133} (\bibinfo {year} {1965})}\BibitemShut {NoStop}%
\bibitem [{\citenamefont {Mori-S\'{a}nchez}\ \emph {et~al.}(2008)\citenamefont
  {Mori-S\'{a}nchez}, \citenamefont {Cohen},\ and\ \citenamefont
  {Yang}}]{Mori-sanchez:2008}%
  \BibitemOpen
  \bibfield  {author} {\bibinfo {author} {\bibfnamefont {P.}~\bibnamefont
  {Mori-S\'{a}nchez}}, \bibinfo {author} {\bibfnamefont {A.~J.}\ \bibnamefont
  {Cohen}}, \ and\ \bibinfo {author} {\bibfnamefont {W.}~\bibnamefont {Yang}},\
  }\href {\doibase 10.1103/PhysRevLett.100.146401} {\bibfield  {journal}
  {\bibinfo  {journal} {Physical Review Letters}\ }\textbf {\bibinfo {volume}
  {100}},\ \bibinfo {pages} {146401} (\bibinfo {year} {2008})}\BibitemShut
  {NoStop}%
\bibitem [{\citenamefont {Schleder}\ \emph {et~al.}(2019)\citenamefont
  {Schleder}, \citenamefont {Padilha}, \citenamefont {Acosta}, \citenamefont
  {Costa},\ and\ \citenamefont {Fazzio}}]{Schleder:2019}%
  \BibitemOpen
  \bibfield  {author} {\bibinfo {author} {\bibfnamefont {G.~R.}\ \bibnamefont
  {Schleder}}, \bibinfo {author} {\bibfnamefont {A.~C.~M.}\ \bibnamefont
  {Padilha}}, \bibinfo {author} {\bibfnamefont {C.~M.}\ \bibnamefont {Acosta}},
  \bibinfo {author} {\bibfnamefont {M.}~\bibnamefont {Costa}}, \ and\ \bibinfo
  {author} {\bibfnamefont {A.}~\bibnamefont {Fazzio}},\ }\href {\doibase
  10.1088/2515-7639/ab084b} {\bibfield  {journal} {\bibinfo  {journal} {Journal
  of Physics: Materials}\ }\textbf {\bibinfo {volume} {2}},\ \bibinfo {pages}
  {032001} (\bibinfo {year} {2019})}\BibitemShut {NoStop}%
\bibitem [{\citenamefont {Lundgaard}\ \emph {et~al.}(2016)\citenamefont
  {Lundgaard}, \citenamefont {Wellendorff}, \citenamefont {Voss}, \citenamefont
  {Jacobsen},\ and\ \citenamefont {Bligaard}}]{Lundgaard:2016}%
  \BibitemOpen
  \bibfield  {author} {\bibinfo {author} {\bibfnamefont {K.~T.}\ \bibnamefont
  {Lundgaard}}, \bibinfo {author} {\bibfnamefont {J.}~\bibnamefont
  {Wellendorff}}, \bibinfo {author} {\bibfnamefont {J.}~\bibnamefont {Voss}},
  \bibinfo {author} {\bibfnamefont {K.~W.}\ \bibnamefont {Jacobsen}}, \ and\
  \bibinfo {author} {\bibfnamefont {T.}~\bibnamefont {Bligaard}},\ }\href
  {\doibase 10.1103/PhysRevB.93.235162} {\bibfield  {journal} {\bibinfo
  {journal} {Physical Review B}\ }\textbf {\bibinfo {volume} {93}},\ \bibinfo
  {pages} {235162} (\bibinfo {year} {2016})}\BibitemShut {NoStop}%
\bibitem [{\citenamefont {Kolb}\ \emph {et~al.}(2017)\citenamefont {Kolb},
  \citenamefont {Lentz},\ and\ \citenamefont {Kolpak}}]{Kolb:2017}%
  \BibitemOpen
  \bibfield  {author} {\bibinfo {author} {\bibfnamefont {B.}~\bibnamefont
  {Kolb}}, \bibinfo {author} {\bibfnamefont {L.~C.}\ \bibnamefont {Lentz}}, \
  and\ \bibinfo {author} {\bibfnamefont {A.~M.}\ \bibnamefont {Kolpak}},\
  }\href {\doibase 10.1038/s41598-017-01251-z} {\bibfield  {journal} {\bibinfo
  {journal} {Scientific Reports}\ }\textbf {\bibinfo {volume} {7}},\ \bibinfo
  {pages} {1192} (\bibinfo {year} {2017})}\BibitemShut {NoStop}%
\bibitem [{\citenamefont {Liu}\ \emph {et~al.}(2017)\citenamefont {Liu},
  \citenamefont {Wang}, \citenamefont {Du}, \citenamefont {Hu}, \citenamefont
  {Zheng},\ and\ \citenamefont {Chen}}]{Liu:2017}%
  \BibitemOpen
  \bibfield  {author} {\bibinfo {author} {\bibfnamefont {Q.}~\bibnamefont
  {Liu}}, \bibinfo {author} {\bibfnamefont {J.}~\bibnamefont {Wang}}, \bibinfo
  {author} {\bibfnamefont {P.}~\bibnamefont {Du}}, \bibinfo {author}
  {\bibfnamefont {L.}~\bibnamefont {Hu}}, \bibinfo {author} {\bibfnamefont
  {X.}~\bibnamefont {Zheng}}, \ and\ \bibinfo {author} {\bibfnamefont
  {G.}~\bibnamefont {Chen}},\ }\href {\doibase 10.1021/acs.jpca.7b07045}
  {\bibfield  {journal} {\bibinfo  {journal} {The Journal of Physical Chemistry
  A}\ }\textbf {\bibinfo {volume} {121}},\ \bibinfo {pages} {7273} (\bibinfo
  {year} {2017})}\BibitemShut {NoStop}%
\bibitem [{\citenamefont {Nagai}\ \emph {et~al.}(2018)\citenamefont {Nagai},
  \citenamefont {Akashi}, \citenamefont {Sasaki},\ and\ \citenamefont
  {Tsuneyuki}}]{Nagai:2018}%
  \BibitemOpen
  \bibfield  {author} {\bibinfo {author} {\bibfnamefont {R.}~\bibnamefont
  {Nagai}}, \bibinfo {author} {\bibfnamefont {R.}~\bibnamefont {Akashi}},
  \bibinfo {author} {\bibfnamefont {S.}~\bibnamefont {Sasaki}}, \ and\ \bibinfo
  {author} {\bibfnamefont {S.}~\bibnamefont {Tsuneyuki}},\ }\href {\doibase
  10.1063/1.5029279} {\bibfield  {journal} {\bibinfo  {journal} {The Journal of
  Chemical Physics}\ }\textbf {\bibinfo {volume} {148}},\ \bibinfo {pages}
  {241737} (\bibinfo {year} {2018})}\BibitemShut {NoStop}%
\bibitem [{\citenamefont {Dick}\ and\ \citenamefont
  {Fernandez-Serra}(2019)}]{Dick:2019}%
  \BibitemOpen
  \bibfield  {author} {\bibinfo {author} {\bibfnamefont {S.}~\bibnamefont
  {Dick}}\ and\ \bibinfo {author} {\bibfnamefont {M.}~\bibnamefont
  {Fernandez-Serra}},\ }\href {\doibase 10.26434/chemrxiv.9947312.v2} {\
  (\bibinfo {year} {2019}),\ 10.26434/chemrxiv.9947312.v2}\BibitemShut
  {NoStop}%
\bibitem [{\citenamefont {Schmidt}\ \emph {et~al.}(2019)\citenamefont
  {Schmidt}, \citenamefont {Benavides-Riveros},\ and\ \citenamefont
  {Marques}}]{Schmidt:2019}%
  \BibitemOpen
  \bibfield  {author} {\bibinfo {author} {\bibfnamefont {J.}~\bibnamefont
  {Schmidt}}, \bibinfo {author} {\bibfnamefont {C.~L.}\ \bibnamefont
  {Benavides-Riveros}}, \ and\ \bibinfo {author} {\bibfnamefont {M.~A.~L.}\
  \bibnamefont {Marques}},\ }\href {\doibase 10.1021/acs.jpclett.9b02422}
  {\bibfield  {journal} {\bibinfo  {journal} {The Journal of Physical Chemistry
  Letters}\ }\textbf {\bibinfo {volume} {10}},\ \bibinfo {pages} {6425}
  (\bibinfo {year} {2019})},\ \bibinfo {note} {arXiv: 1908.06198}\BibitemShut
  {NoStop}%
\bibitem [{\citenamefont {Snyder}\ \emph {et~al.}(2012)\citenamefont {Snyder},
  \citenamefont {Rupp}, \citenamefont {Hansen}, \citenamefont {M{\"u}ller},\
  and\ \citenamefont {Burke}}]{Snyder:2012}%
  \BibitemOpen
  \bibfield  {author} {\bibinfo {author} {\bibfnamefont {J.~C.}\ \bibnamefont
  {Snyder}}, \bibinfo {author} {\bibfnamefont {M.}~\bibnamefont {Rupp}},
  \bibinfo {author} {\bibfnamefont {K.}~\bibnamefont {Hansen}}, \bibinfo
  {author} {\bibfnamefont {K.-R.}\ \bibnamefont {M{\"u}ller}}, \ and\ \bibinfo
  {author} {\bibfnamefont {K.}~\bibnamefont {Burke}},\ }\href {\doibase
  10.1103/PhysRevLett.108.253002} {\bibfield  {journal} {\bibinfo  {journal}
  {Physical Review Letters}\ }\textbf {\bibinfo {volume} {108}},\ \bibinfo
  {pages} {253002} (\bibinfo {year} {2012})}\BibitemShut {NoStop}%
\bibitem [{\citenamefont {Snyder}\ \emph {et~al.}(2013)\citenamefont {Snyder},
  \citenamefont {Rupp}, \citenamefont {Hansen}, \citenamefont {Blooston},
  \citenamefont {M{\"u}ller},\ and\ \citenamefont {Burke}}]{Snyder:2013}%
  \BibitemOpen
  \bibfield  {author} {\bibinfo {author} {\bibfnamefont {J.~C.}\ \bibnamefont
  {Snyder}}, \bibinfo {author} {\bibfnamefont {M.}~\bibnamefont {Rupp}},
  \bibinfo {author} {\bibfnamefont {K.}~\bibnamefont {Hansen}}, \bibinfo
  {author} {\bibfnamefont {L.}~\bibnamefont {Blooston}}, \bibinfo {author}
  {\bibfnamefont {K.-R.}\ \bibnamefont {M{\"u}ller}}, \ and\ \bibinfo {author}
  {\bibfnamefont {K.}~\bibnamefont {Burke}},\ }\href {\doibase
  10.1063/1.4834075} {\bibfield  {journal} {\bibinfo  {journal} {The Journal of
  Chemical Physics}\ }\textbf {\bibinfo {volume} {139}},\ \bibinfo {pages}
  {224104} (\bibinfo {year} {2013})}\BibitemShut {NoStop}%
\bibitem [{\citenamefont {Li}\ \emph {et~al.}(2016{\natexlab{a}})\citenamefont
  {Li}, \citenamefont {Snyder}, \citenamefont {Pelaschier}, \citenamefont
  {Huang}, \citenamefont {Niranjan}, \citenamefont {Duncan}, \citenamefont
  {Rupp}, \citenamefont {M{\"u}ller},\ and\ \citenamefont {Burke}}]{Li:2016a}%
  \BibitemOpen
  \bibfield  {author} {\bibinfo {author} {\bibfnamefont {L.}~\bibnamefont
  {Li}}, \bibinfo {author} {\bibfnamefont {J.~C.}\ \bibnamefont {Snyder}},
  \bibinfo {author} {\bibfnamefont {I.~M.}\ \bibnamefont {Pelaschier}},
  \bibinfo {author} {\bibfnamefont {J.}~\bibnamefont {Huang}}, \bibinfo
  {author} {\bibfnamefont {U.-N.}\ \bibnamefont {Niranjan}}, \bibinfo {author}
  {\bibfnamefont {P.}~\bibnamefont {Duncan}}, \bibinfo {author} {\bibfnamefont
  {M.}~\bibnamefont {Rupp}}, \bibinfo {author} {\bibfnamefont {K.-R.}\
  \bibnamefont {M{\"u}ller}}, \ and\ \bibinfo {author} {\bibfnamefont
  {K.}~\bibnamefont {Burke}},\ }\href {\doibase 10.1002/qua.25040} {\bibfield
  {journal} {\bibinfo  {journal} {International Journal of Quantum Chemistry}\
  }\textbf {\bibinfo {volume} {116}},\ \bibinfo {pages} {819} (\bibinfo {year}
  {2016}{\natexlab{a}})}\BibitemShut {NoStop}%
\bibitem [{\citenamefont {Li}\ \emph {et~al.}(2016{\natexlab{b}})\citenamefont
  {Li}, \citenamefont {Baker}, \citenamefont {White},\ and\ \citenamefont
  {Burke}}]{Li:2016b}%
  \BibitemOpen
  \bibfield  {author} {\bibinfo {author} {\bibfnamefont {L.}~\bibnamefont
  {Li}}, \bibinfo {author} {\bibfnamefont {T.~E.}\ \bibnamefont {Baker}},
  \bibinfo {author} {\bibfnamefont {S.~R.}\ \bibnamefont {White}}, \ and\
  \bibinfo {author} {\bibfnamefont {K.}~\bibnamefont {Burke}},\ }\href
  {\doibase 10.1103/PhysRevB.94.245129} {\bibfield  {journal} {\bibinfo
  {journal} {Physical Review B}\ }\textbf {\bibinfo {volume} {94}},\ \bibinfo
  {pages} {245129} (\bibinfo {year} {2016}{\natexlab{b}})}\BibitemShut
  {NoStop}%
\bibitem [{\citenamefont {Yao}\ and\ \citenamefont
  {Parkhill}(2016)}]{Yao:2016}%
  \BibitemOpen
  \bibfield  {author} {\bibinfo {author} {\bibfnamefont {K.}~\bibnamefont
  {Yao}}\ and\ \bibinfo {author} {\bibfnamefont {J.}~\bibnamefont {Parkhill}},\
  }\href {\doibase 10.1021/acs.jctc.5b01011} {\bibfield  {journal} {\bibinfo
  {journal} {Journal of Chemical Theory and Computation}\ }\textbf {\bibinfo
  {volume} {12}},\ \bibinfo {pages} {1139} (\bibinfo {year}
  {2016})}\BibitemShut {NoStop}%
\bibitem [{\citenamefont {Seino}\ \emph {et~al.}(2018)\citenamefont {Seino},
  \citenamefont {Kageyama}, \citenamefont {Fujinami}, \citenamefont {Ikabata},\
  and\ \citenamefont {Nakai}}]{Seino:2018}%
  \BibitemOpen
  \bibfield  {author} {\bibinfo {author} {\bibfnamefont {J.}~\bibnamefont
  {Seino}}, \bibinfo {author} {\bibfnamefont {R.}~\bibnamefont {Kageyama}},
  \bibinfo {author} {\bibfnamefont {M.}~\bibnamefont {Fujinami}}, \bibinfo
  {author} {\bibfnamefont {Y.}~\bibnamefont {Ikabata}}, \ and\ \bibinfo
  {author} {\bibfnamefont {H.}~\bibnamefont {Nakai}},\ }\href {\doibase
  10.1063/1.5007230} {\bibfield  {journal} {\bibinfo  {journal} {The Journal of
  Chemical Physics}\ }\textbf {\bibinfo {volume} {148}},\ \bibinfo {pages}
  {241705} (\bibinfo {year} {2018})}\BibitemShut {NoStop}%
\bibitem [{\citenamefont {Nelson}\ \emph {et~al.}(2019)\citenamefont {Nelson},
  \citenamefont {Tiwari},\ and\ \citenamefont {Sanvito}}]{Nelson:2019}%
  \BibitemOpen
  \bibfield  {author} {\bibinfo {author} {\bibfnamefont {J.}~\bibnamefont
  {Nelson}}, \bibinfo {author} {\bibfnamefont {R.}~\bibnamefont {Tiwari}}, \
  and\ \bibinfo {author} {\bibfnamefont {S.}~\bibnamefont {Sanvito}},\ }\href
  {\doibase 10.1103/PhysRevB.99.075132} {\bibfield  {journal} {\bibinfo
  {journal} {Physical Review B}\ }\textbf {\bibinfo {volume} {99}},\ \bibinfo
  {pages} {075132} (\bibinfo {year} {2019})}\BibitemShut {NoStop}%
\bibitem [{\citenamefont {Golub}\ and\ \citenamefont
  {Manzhos}(2019)}]{Golub:2019}%
  \BibitemOpen
  \bibfield  {author} {\bibinfo {author} {\bibfnamefont {P.}~\bibnamefont
  {Golub}}\ and\ \bibinfo {author} {\bibfnamefont {S.}~\bibnamefont
  {Manzhos}},\ }\href {\doibase 10.1039/C8CP06433D} {\bibfield  {journal}
  {\bibinfo  {journal} {Physical Chemistry Chemical Physics}\ }\textbf
  {\bibinfo {volume} {21}},\ \bibinfo {pages} {378} (\bibinfo {year}
  {2019})}\BibitemShut {NoStop}%
\bibitem [{\citenamefont {Nudejima}\ \emph {et~al.}(2019)\citenamefont
  {Nudejima}, \citenamefont {Ikabata}, \citenamefont {Seino}, \citenamefont
  {Yoshikawa},\ and\ \citenamefont {Nakai}}]{Nudejima:2019}%
  \BibitemOpen
  \bibfield  {author} {\bibinfo {author} {\bibfnamefont {T.}~\bibnamefont
  {Nudejima}}, \bibinfo {author} {\bibfnamefont {Y.}~\bibnamefont {Ikabata}},
  \bibinfo {author} {\bibfnamefont {J.}~\bibnamefont {Seino}}, \bibinfo
  {author} {\bibfnamefont {T.}~\bibnamefont {Yoshikawa}}, \ and\ \bibinfo
  {author} {\bibfnamefont {H.}~\bibnamefont {Nakai}},\ }\href {\doibase
  10.1063/1.5100165} {\bibfield  {journal} {\bibinfo  {journal} {The Journal of
  Chemical Physics}\ }\textbf {\bibinfo {volume} {151}},\ \bibinfo {pages}
  {024104} (\bibinfo {year} {2019})}\BibitemShut {NoStop}%
\bibitem [{\citenamefont {Hansen}\ \emph {et~al.}(2013)\citenamefont {Hansen},
  \citenamefont {Montavon}, \citenamefont {Biegler}, \citenamefont {Fazli},
  \citenamefont {Rupp}, \citenamefont {Scheffler}, \citenamefont {von
  Lilienfeld}, \citenamefont {Tkatchenko},\ and\ \citenamefont
  {M{\"u}ller}}]{Hansen:2013}%
  \BibitemOpen
  \bibfield  {author} {\bibinfo {author} {\bibfnamefont {K.}~\bibnamefont
  {Hansen}}, \bibinfo {author} {\bibfnamefont {G.}~\bibnamefont {Montavon}},
  \bibinfo {author} {\bibfnamefont {F.}~\bibnamefont {Biegler}}, \bibinfo
  {author} {\bibfnamefont {S.}~\bibnamefont {Fazli}}, \bibinfo {author}
  {\bibfnamefont {M.}~\bibnamefont {Rupp}}, \bibinfo {author} {\bibfnamefont
  {M.}~\bibnamefont {Scheffler}}, \bibinfo {author} {\bibfnamefont {O.~A.}\
  \bibnamefont {von Lilienfeld}}, \bibinfo {author} {\bibfnamefont
  {A.}~\bibnamefont {Tkatchenko}}, \ and\ \bibinfo {author} {\bibfnamefont
  {K.-R.}\ \bibnamefont {M{\"u}ller}},\ }\href {\doibase 10.1021/ct400195d}
  {\bibfield  {journal} {\bibinfo  {journal} {Journal of Chemical Theory and
  Computation}\ }\textbf {\bibinfo {volume} {9}},\ \bibinfo {pages} {3404}
  (\bibinfo {year} {2013})}\BibitemShut {NoStop}%
\bibitem [{\citenamefont {Sch{\"u}tt}\ \emph {et~al.}(2014)\citenamefont
  {Sch{\"u}tt}, \citenamefont {Glawe}, \citenamefont {Brockherde},
  \citenamefont {Sanna}, \citenamefont {M{\"u}ller},\ and\ \citenamefont
  {Gross}}]{Schutt:2014}%
  \BibitemOpen
  \bibfield  {author} {\bibinfo {author} {\bibfnamefont {K.~T.}\ \bibnamefont
  {Sch{\"u}tt}}, \bibinfo {author} {\bibfnamefont {H.}~\bibnamefont {Glawe}},
  \bibinfo {author} {\bibfnamefont {F.}~\bibnamefont {Brockherde}}, \bibinfo
  {author} {\bibfnamefont {A.}~\bibnamefont {Sanna}}, \bibinfo {author}
  {\bibfnamefont {K.~R.}\ \bibnamefont {M{\"u}ller}}, \ and\ \bibinfo {author}
  {\bibfnamefont {E.~K.~U.}\ \bibnamefont {Gross}},\ }\href {\doibase
  10.1103/PhysRevB.89.205118} {\bibfield  {journal} {\bibinfo  {journal}
  {Physical Review B}\ }\textbf {\bibinfo {volume} {89}},\ \bibinfo {pages}
  {205118} (\bibinfo {year} {2014})}\BibitemShut {NoStop}%
\bibitem [{\citenamefont {Hansen}\ \emph {et~al.}(2015)\citenamefont {Hansen},
  \citenamefont {Biegler}, \citenamefont {Ramakrishnan}, \citenamefont
  {Pronobis}, \citenamefont {von Lilienfeld}, \citenamefont {M{\"u}ller},\ and\
  \citenamefont {Tkatchenko}}]{Hansen:2015}%
  \BibitemOpen
  \bibfield  {author} {\bibinfo {author} {\bibfnamefont {K.}~\bibnamefont
  {Hansen}}, \bibinfo {author} {\bibfnamefont {F.}~\bibnamefont {Biegler}},
  \bibinfo {author} {\bibfnamefont {R.}~\bibnamefont {Ramakrishnan}}, \bibinfo
  {author} {\bibfnamefont {W.}~\bibnamefont {Pronobis}}, \bibinfo {author}
  {\bibfnamefont {O.~A.}\ \bibnamefont {von Lilienfeld}}, \bibinfo {author}
  {\bibfnamefont {K.-R.}\ \bibnamefont {M{\"u}ller}}, \ and\ \bibinfo {author}
  {\bibfnamefont {A.}~\bibnamefont {Tkatchenko}},\ }\href {\doibase
  10.1021/acs.jpclett.5b00831} {\bibfield  {journal} {\bibinfo  {journal} {The
  Journal of Physical Chemistry Letters}\ }\textbf {\bibinfo {volume} {6}},\
  \bibinfo {pages} {2326} (\bibinfo {year} {2015})}\BibitemShut {NoStop}%
\bibitem [{\citenamefont {Sch{\"u}tt}\ \emph {et~al.}(2017)\citenamefont
  {Sch{\"u}tt}, \citenamefont {Arbabzadah}, \citenamefont {Chmiela},
  \citenamefont {M{\"u}ller},\ and\ \citenamefont {Tkatchenko}}]{Schutt:2017}%
  \BibitemOpen
  \bibfield  {author} {\bibinfo {author} {\bibfnamefont {K.~T.}\ \bibnamefont
  {Sch{\"u}tt}}, \bibinfo {author} {\bibfnamefont {F.}~\bibnamefont
  {Arbabzadah}}, \bibinfo {author} {\bibfnamefont {S.}~\bibnamefont {Chmiela}},
  \bibinfo {author} {\bibfnamefont {K.~R.}\ \bibnamefont {M{\"u}ller}}, \ and\
  \bibinfo {author} {\bibfnamefont {A.}~\bibnamefont {Tkatchenko}},\ }\href
  {\doibase 10.1038/ncomms13890} {\bibfield  {journal} {\bibinfo  {journal}
  {Nature Communications}\ }\textbf {\bibinfo {volume} {8}},\ \bibinfo {pages}
  {13890} (\bibinfo {year} {2017})}\BibitemShut {NoStop}%
\bibitem [{\citenamefont {Brockherde}\ \emph {et~al.}(2017)\citenamefont
  {Brockherde}, \citenamefont {Vogt}, \citenamefont {Li}, \citenamefont
  {Tuckerman}, \citenamefont {Burke},\ and\ \citenamefont
  {M{\"u}ller}}]{Brockherde:2017}%
  \BibitemOpen
  \bibfield  {author} {\bibinfo {author} {\bibfnamefont {F.}~\bibnamefont
  {Brockherde}}, \bibinfo {author} {\bibfnamefont {L.}~\bibnamefont {Vogt}},
  \bibinfo {author} {\bibfnamefont {L.}~\bibnamefont {Li}}, \bibinfo {author}
  {\bibfnamefont {M.~E.}\ \bibnamefont {Tuckerman}}, \bibinfo {author}
  {\bibfnamefont {K.}~\bibnamefont {Burke}}, \ and\ \bibinfo {author}
  {\bibfnamefont {K.-R.}\ \bibnamefont {M{\"u}ller}},\ }\href {\doibase
  10.1038/s41467-017-00839-3} {\bibfield  {journal} {\bibinfo  {journal}
  {Nature Communications}\ }\textbf {\bibinfo {volume} {8}},\ \bibinfo {pages}
  {872} (\bibinfo {year} {2017})}\BibitemShut {NoStop}%
\bibitem [{\citenamefont {{Bogojeski}}\ \emph {et~al.}(2018)\citenamefont
  {{Bogojeski}}, \citenamefont {{Brockherde}}, \citenamefont {{Vogt-Maranto}},
  \citenamefont {{Li}}, \citenamefont {{Tuckerman}}, \citenamefont {{Burke}},\
  and\ \citenamefont {{M{\"u}ller}}}]{Bogojeski:2018}%
  \BibitemOpen
  \bibfield  {author} {\bibinfo {author} {\bibfnamefont {M.}~\bibnamefont
  {{Bogojeski}}}, \bibinfo {author} {\bibfnamefont {F.}~\bibnamefont
  {{Brockherde}}}, \bibinfo {author} {\bibfnamefont {L.}~\bibnamefont
  {{Vogt-Maranto}}}, \bibinfo {author} {\bibfnamefont {L.}~\bibnamefont
  {{Li}}}, \bibinfo {author} {\bibfnamefont {M.~E.}\ \bibnamefont
  {{Tuckerman}}}, \bibinfo {author} {\bibfnamefont {K.}~\bibnamefont
  {{Burke}}}, \ and\ \bibinfo {author} {\bibfnamefont {K.-R.}\ \bibnamefont
  {{M{\"u}ller}}},\ }\href@noop {} {\bibfield  {journal} {\bibinfo  {journal}
  {arXiv:1811.06255 [physics.comp-ph]}\ } (\bibinfo {year} {2018})},\ \bibinfo
  {note} {arXiv:1811.06255}\BibitemShut {NoStop}%
\bibitem [{\citenamefont {Ryczko}\ \emph {et~al.}(2018)\citenamefont {Ryczko},
  \citenamefont {Mills}, \citenamefont {Luchak}, \citenamefont {Homenick},\
  and\ \citenamefont {Tamblyn}}]{Ryczko:2018}%
  \BibitemOpen
  \bibfield  {author} {\bibinfo {author} {\bibfnamefont {K.}~\bibnamefont
  {Ryczko}}, \bibinfo {author} {\bibfnamefont {K.}~\bibnamefont {Mills}},
  \bibinfo {author} {\bibfnamefont {I.}~\bibnamefont {Luchak}}, \bibinfo
  {author} {\bibfnamefont {C.}~\bibnamefont {Homenick}}, \ and\ \bibinfo
  {author} {\bibfnamefont {I.}~\bibnamefont {Tamblyn}},\ }\href {\doibase
  10.1016/j.commatsci.2018.03.005} {\bibfield  {journal} {\bibinfo  {journal}
  {Computational Materials Science}\ }\textbf {\bibinfo {volume} {149}},\
  \bibinfo {pages} {134} (\bibinfo {year} {2018})}\BibitemShut {NoStop}%
\bibitem [{\citenamefont {Schmidt}\ \emph {et~al.}(2018)\citenamefont
  {Schmidt}, \citenamefont {Fowler}, \citenamefont {Elliott},\ and\
  \citenamefont {Bristowe}}]{Schmidt:2018}%
  \BibitemOpen
  \bibfield  {author} {\bibinfo {author} {\bibfnamefont {E.}~\bibnamefont
  {Schmidt}}, \bibinfo {author} {\bibfnamefont {A.~T.}\ \bibnamefont {Fowler}},
  \bibinfo {author} {\bibfnamefont {J.~A.}\ \bibnamefont {Elliott}}, \ and\
  \bibinfo {author} {\bibfnamefont {P.~D.}\ \bibnamefont {Bristowe}},\ }\href
  {\doibase https://doi.org/10.1016/j.commatsci.2018.03.029} {\bibfield
  {journal} {\bibinfo  {journal} {Computational Materials Science}\ }\textbf
  {\bibinfo {volume} {149}},\ \bibinfo {pages} {250 } (\bibinfo {year}
  {2018})}\BibitemShut {NoStop}%
\bibitem [{\citenamefont {Pilati}\ and\ \citenamefont
  {Pieri}(2019)}]{Pilati:2019}%
  \BibitemOpen
  \bibfield  {author} {\bibinfo {author} {\bibfnamefont {S.}~\bibnamefont
  {Pilati}}\ and\ \bibinfo {author} {\bibfnamefont {P.}~\bibnamefont {Pieri}},\
  }\href {\doibase 10.1038/s41598-019-42125-w} {\bibfield  {journal} {\bibinfo
  {journal} {Scientific Reports}\ }\textbf {\bibinfo {volume} {9}},\ \bibinfo
  {pages} {5613} (\bibinfo {year} {2019})}\BibitemShut {NoStop}%
\bibitem [{\citenamefont {Custodio}\ \emph {et~al.}(2019)\citenamefont
  {Custodio}, \citenamefont {Filletti},\ and\ \citenamefont
  {Fran\c{c}a}}]{Custodio:2019}%
  \BibitemOpen
  \bibfield  {author} {\bibinfo {author} {\bibfnamefont {C.~A.}\ \bibnamefont
  {Custodio}}, \bibinfo {author} {\bibfnamefont {E.~R.}\ \bibnamefont
  {Filletti}}, \ and\ \bibinfo {author} {\bibfnamefont {V.~V.}\ \bibnamefont
  {Fran\c{c}a}},\ }\href {\doibase 10.1038/s41598-018-37999-1} {\bibfield
  {journal} {\bibinfo  {journal} {Scientific Reports}\ }\textbf {\bibinfo
  {volume} {9}},\ \bibinfo {pages} {1886} (\bibinfo {year} {2019})},\ \bibinfo
  {note} {arXiv: 1811.03774}\BibitemShut {NoStop}%
\bibitem [{\citenamefont {Ryczko}\ \emph {et~al.}(2019)\citenamefont {Ryczko},
  \citenamefont {Strubbe},\ and\ \citenamefont {Tamblyn}}]{Ryczko:2019}%
  \BibitemOpen
  \bibfield  {author} {\bibinfo {author} {\bibfnamefont {K.}~\bibnamefont
  {Ryczko}}, \bibinfo {author} {\bibfnamefont {D.~A.}\ \bibnamefont {Strubbe}},
  \ and\ \bibinfo {author} {\bibfnamefont {I.}~\bibnamefont {Tamblyn}},\ }\href
  {\doibase 10.1103/PhysRevA.100.022512} {\bibfield  {journal} {\bibinfo
  {journal} {Physical Review A}\ }\textbf {\bibinfo {volume} {100}},\ \bibinfo
  {pages} {022512} (\bibinfo {year} {2019})}\BibitemShut {NoStop}%
\bibitem [{\citenamefont {{Zepeda-N{\'u}{\~n}ez}}\ \emph
  {et~al.}(2019)\citenamefont {{Zepeda-N{\'u}{\~n}ez}}, \citenamefont {{Chen}},
  \citenamefont {{Zhang}}, \citenamefont {{Jia}}, \citenamefont {{Zhang}},\
  and\ \citenamefont {{Lin}}}]{Zepeda:2019}%
  \BibitemOpen
  \bibfield  {author} {\bibinfo {author} {\bibfnamefont {L.}~\bibnamefont
  {{Zepeda-N{\'u}{\~n}ez}}}, \bibinfo {author} {\bibfnamefont {Y.}~\bibnamefont
  {{Chen}}}, \bibinfo {author} {\bibfnamefont {J.}~\bibnamefont {{Zhang}}},
  \bibinfo {author} {\bibfnamefont {W.}~\bibnamefont {{Jia}}}, \bibinfo
  {author} {\bibfnamefont {L.}~\bibnamefont {{Zhang}}}, \ and\ \bibinfo
  {author} {\bibfnamefont {L.}~\bibnamefont {{Lin}}},\ }\href@noop {}
  {\bibfield  {journal} {\bibinfo  {journal} {arXiv:1912.00775
  [physics.comp-ph]}\ } (\bibinfo {year} {2019})},\ \bibinfo {note}
  {arXiv:1912.00775}\BibitemShut {NoStop}%
\bibitem [{\citenamefont {Moreno}\ \emph {et~al.}(2019)\citenamefont {Moreno},
  \citenamefont {Carleo},\ and\ \citenamefont {Georges}}]{Moreno:2019}%
  \BibitemOpen
  \bibfield  {author} {\bibinfo {author} {\bibfnamefont {J.~R.}\ \bibnamefont
  {Moreno}}, \bibinfo {author} {\bibfnamefont {G.}~\bibnamefont {Carleo}}, \
  and\ \bibinfo {author} {\bibfnamefont {A.}~\bibnamefont {Georges}},\ }\href
  {http://arxiv.org/abs/1911.03580} {\bibfield  {journal} {\bibinfo  {journal}
  {arXiv:1911.03580 [cond-mat, physics:quant-ph]}\ } (\bibinfo {year}
  {2019})},\ \bibinfo {note} {arXiv: 1911.03580}\BibitemShut {NoStop}%
\bibitem [{\citenamefont {Settles}(2009)}]{Settles:2009}%
  \BibitemOpen
  \bibfield  {author} {\bibinfo {author} {\bibfnamefont {B.}~\bibnamefont
  {Settles}},\ }\href@noop {} {\emph {\bibinfo {title} {Active Learning
  Literature Survey}}},\ \bibinfo {type} {Computer Sciences Technical Report}\
  \bibinfo {number} {1648}\ (\bibinfo  {institution} {University of
  Wisconsin--Madison},\ \bibinfo {year} {2009})\BibitemShut {NoStop}%
\bibitem [{\citenamefont {Gubaev}\ \emph {et~al.}(2019)\citenamefont {Gubaev},
  \citenamefont {Podryabinkin}, \citenamefont {Hart},\ and\ \citenamefont
  {Shapeev}}]{Gubaev:2019}%
  \BibitemOpen
  \bibfield  {author} {\bibinfo {author} {\bibfnamefont {K.}~\bibnamefont
  {Gubaev}}, \bibinfo {author} {\bibfnamefont {E.~V.}\ \bibnamefont
  {Podryabinkin}}, \bibinfo {author} {\bibfnamefont {G.~L.}\ \bibnamefont
  {Hart}}, \ and\ \bibinfo {author} {\bibfnamefont {A.~V.}\ \bibnamefont
  {Shapeev}},\ }\href {\doibase
  https://doi.org/10.1016/j.commatsci.2018.09.031} {\bibfield  {journal}
  {\bibinfo  {journal} {Computational Materials Science}\ }\textbf {\bibinfo
  {volume} {156}},\ \bibinfo {pages} {148 } (\bibinfo {year}
  {2019})}\BibitemShut {NoStop}%
\bibitem [{\citenamefont {{Sivaraman}}\ \emph {et~al.}(2019)\citenamefont
  {{Sivaraman}}, \citenamefont {{Narayanan Krishnamoorthy}}, \citenamefont
  {{Baur}}, \citenamefont {{Holm}}, \citenamefont {{Stan}}, \citenamefont
  {{Cs{\'a}nyi}}, \citenamefont {{Benmore}},\ and\ \citenamefont
  {{V{\'a}zquez-Mayagoitia}}}]{Sivaraman:2019}%
  \BibitemOpen
  \bibfield  {author} {\bibinfo {author} {\bibfnamefont {G.}~\bibnamefont
  {{Sivaraman}}}, \bibinfo {author} {\bibfnamefont {A.}~\bibnamefont
  {{Narayanan Krishnamoorthy}}}, \bibinfo {author} {\bibfnamefont
  {M.}~\bibnamefont {{Baur}}}, \bibinfo {author} {\bibfnamefont
  {C.}~\bibnamefont {{Holm}}}, \bibinfo {author} {\bibfnamefont
  {M.}~\bibnamefont {{Stan}}}, \bibinfo {author} {\bibfnamefont
  {G.}~\bibnamefont {{Cs{\'a}nyi}}}, \bibinfo {author} {\bibfnamefont
  {C.}~\bibnamefont {{Benmore}}}, \ and\ \bibinfo {author} {\bibfnamefont
  {{\'A}.}~\bibnamefont {{V{\'a}zquez-Mayagoitia}}},\ }\href@noop {} {\bibfield
   {journal} {\bibinfo  {journal} {arXiv:1910.10254 [cond-mat.mtrl-sci]}\ }
  (\bibinfo {year} {2019})},\ \bibinfo {note} {arXiv:1910.10254}\BibitemShut
  {NoStop}%
\bibitem [{\citenamefont {Yao}\ \emph {et~al.}(2020)\citenamefont {Yao},
  \citenamefont {Wu}, \citenamefont {Koo}, \citenamefont {Yan},\ and\
  \citenamefont {Zhai}}]{Yao:2020}%
  \BibitemOpen
  \bibfield  {author} {\bibinfo {author} {\bibfnamefont {J.}~\bibnamefont
  {Yao}}, \bibinfo {author} {\bibfnamefont {Y.}~\bibnamefont {Wu}}, \bibinfo
  {author} {\bibfnamefont {J.}~\bibnamefont {Koo}}, \bibinfo {author}
  {\bibfnamefont {B.}~\bibnamefont {Yan}}, \ and\ \bibinfo {author}
  {\bibfnamefont {H.}~\bibnamefont {Zhai}},\ }\href {\doibase
  10.1103/PhysRevResearch.2.013287} {\bibfield  {journal} {\bibinfo  {journal}
  {Phys. Rev. Research}\ }\textbf {\bibinfo {volume} {2}},\ \bibinfo {pages}
  {013287} (\bibinfo {year} {2020})}\BibitemShut {NoStop}%
\bibitem [{\citenamefont {{Teichert}}\ \emph {et~al.}(2020)\citenamefont
  {{Teichert}}, \citenamefont {{Natarajan}}, \citenamefont {{Van der Ven}},\
  and\ \citenamefont {{Garikipati}}}]{Teichert:2020}%
  \BibitemOpen
  \bibfield  {author} {\bibinfo {author} {\bibfnamefont {G.}~\bibnamefont
  {{Teichert}}}, \bibinfo {author} {\bibfnamefont {A.}~\bibnamefont
  {{Natarajan}}}, \bibinfo {author} {\bibfnamefont {A.}~\bibnamefont {{Van der
  Ven}}}, \ and\ \bibinfo {author} {\bibfnamefont {K.}~\bibnamefont
  {{Garikipati}}},\ }\href@noop {} {\bibfield  {journal} {\bibinfo  {journal}
  {arXiv:2002.02305 [cs.LG]}\ } (\bibinfo {year} {2020})},\ \bibinfo {note}
  {arXiv:2002.02305}\BibitemShut {NoStop}%
\bibitem [{\citenamefont {Sch\"onhammer}\ \emph {et~al.}(1995)\citenamefont
  {Sch\"onhammer}, \citenamefont {Gunnarsson},\ and\ \citenamefont
  {Noack}}]{Schonhammer:1995}%
  \BibitemOpen
  \bibfield  {author} {\bibinfo {author} {\bibfnamefont {K.}~\bibnamefont
  {Sch\"onhammer}}, \bibinfo {author} {\bibfnamefont {O.}~\bibnamefont
  {Gunnarsson}}, \ and\ \bibinfo {author} {\bibfnamefont {R.~M.}\ \bibnamefont
  {Noack}},\ }\href {\doibase 10.1103/PhysRevB.52.2504} {\bibfield  {journal}
  {\bibinfo  {journal} {Phys. Rev. B}\ }\textbf {\bibinfo {volume} {52}},\
  \bibinfo {pages} {2504} (\bibinfo {year} {1995})}\BibitemShut {NoStop}%
\bibitem [{\citenamefont {Markhof}\ \emph {et~al.}(2018)\citenamefont
  {Markhof}, \citenamefont {Sbierski}, \citenamefont {Meden},\ and\
  \citenamefont {Karrasch}}]{Markhof:2018}%
  \BibitemOpen
  \bibfield  {author} {\bibinfo {author} {\bibfnamefont {L.}~\bibnamefont
  {Markhof}}, \bibinfo {author} {\bibfnamefont {B.}~\bibnamefont {Sbierski}},
  \bibinfo {author} {\bibfnamefont {V.}~\bibnamefont {Meden}}, \ and\ \bibinfo
  {author} {\bibfnamefont {C.}~\bibnamefont {Karrasch}},\ }\href {\doibase
  10.1103/PhysRevB.97.235126} {\bibfield  {journal} {\bibinfo  {journal}
  {Physical Review B}\ }\textbf {\bibinfo {volume} {97}},\ \bibinfo {pages}
  {235126} (\bibinfo {year} {2018})}\BibitemShut {NoStop}%
\bibitem [{Note1()}]{Note1}%
  \BibitemOpen
  \bibinfo {note} {In order to also use small systems ($N_{\protect \mathrm
  {uc}}=4$) with periodic boundary conditions as part of the training data, at
  most two adjacent unit cells from the interior of the density-density
  correlator are a valid representative.}\BibitemShut {Stop}%
\bibitem [{\citenamefont {Chollet}\ \emph {et~al.}(2015)\citenamefont {Chollet}
  \emph {et~al.}}]{Chollet:2015}%
  \BibitemOpen
  \bibfield  {author} {\bibinfo {author} {\bibfnamefont {F.}~\bibnamefont
  {Chollet}} \emph {et~al.},\ }\href@noop {} {\enquote {\bibinfo {title}
  {Keras},}\ }\bibinfo {howpublished} {\url{https://keras.io}} (\bibinfo {year}
  {2015})\BibitemShut {NoStop}%
\bibitem [{Note2()}]{Note2}%
  \BibitemOpen
  \bibinfo {note} {Calculations were performed using the TeNPy Library (version
  0.5.0), for a finite but periodic lattice with maximum bond dimension $\chi
  _{\protect \mathrm {max}} = 2000$, max. truncation error $\epsilon =
  10\protect \mathrm {e}{-8}$ and energy convergence criterion $\Delta E =
  10\protect \mathrm {e}{-6}$.}\BibitemShut {Stop}%
\bibitem [{\citenamefont {Hauschild}\ and\ \citenamefont
  {Pollmann}(2018)}]{Hauschild:2018}%
  \BibitemOpen
  \bibfield  {author} {\bibinfo {author} {\bibfnamefont {J.}~\bibnamefont
  {Hauschild}}\ and\ \bibinfo {author} {\bibfnamefont {F.}~\bibnamefont
  {Pollmann}},\ }\href {\doibase 10.21468/SciPostPhysLectNotes.5} {\bibfield
  {journal} {\bibinfo  {journal} {SciPost Phys. Lect. Notes 5}\ } (\bibinfo
  {year} {2018}),\ 10.21468/SciPostPhysLectNotes.5},\ \bibinfo {note} {code
  available from \url{https://github.com/tenpy/tenpy}},\ \Eprint
  {http://arxiv.org/abs/1805.00055} {arXiv:1805.00055} \BibitemShut {NoStop}%
\bibitem [{Note3()}]{Note3}%
  \BibitemOpen
  \bibinfo {note} {Note that we choose $C$ to highlight that two neighboring
  sites are always occupied by just one electron. That distinguishes it from
  the band insulator case of filling $n_{\protect \mathrm {e}} = 2$ and a
  spontaneously translation symmetry broken phase at $n_{\protect \mathrm {e}}
  = 1$ (both $C = 0$).}\BibitemShut {Stop}%
\bibitem [{Note4()}]{Note4}%
  \BibitemOpen
  \bibinfo {note} {Calculations were performed using the TeNPy Library (version
  0.5.0), for a finite lattice with maximum bond dimension $\chi _{\protect
  \mathrm {max}} = 1000$, max. truncation error $\epsilon = 10\protect \mathrm
  {e}{-10}$ and energy convergence criterion $\Delta E = 10\protect \mathrm
  {e}{-6}$.}\BibitemShut {Stop}%
\end{thebibliography}%

\end{document}